\documentclass[useAMS]{mn2e}
\usepackage{amsmath}
\usepackage{url}
\usepackage{amsfonts}
\usepackage{amsbsy}
\usepackage{subfigure}
\usepackage{verbatim}
\usepackage{amssymb}
\usepackage{amsmath}
\usepackage{amsbsy}
\usepackage{graphicx}
\usepackage{array}
\usepackage[usenames, dvipsnames]{color}

\title{Turbulent fluctuations and the excitation of Z Cam outbursts}
\author[Ross \& Latter]{Johnathan Ross$^{1}$\thanks{E-mail:
   jpjr2@cam.ac.uk},
   Henrik N. Latter$^{1}$\\
$^{1}$ DAMTP, University of Cambridge, CMS, Wilberforce Road
Cambridge CB3 0WA, UK\\
}
\date{}

\begin{document}

\maketitle

\begin{abstract}
Z Cam variables are a subclass of dwarf nova
that lie near a global bifurcation between outbursting (`limit cycle')
and
non-outbursting (`standstill') states.
It is believed that variations in the secondary star's mass-injection rate
instigate transitions between the two regimes. In this paper
we explore an alternative trigger for these transitions: 
stochastic fluctuations in the disk's turbulent viscosity.
We employ
simple
one-zone and global viscous models which, though inappropriate for
detailed matching to observed light curves, clearly indicate that
turbulent disk fluctuations induce outbursts when the
system is sufficiently close to the global bifurcation point. 
While the models easily produce the observed `outburst/dip' pairs
exhibited by Z Cam and Nova-like variables,
they struggle to generate long trains of outbursts. We conclude that mass
transfer variability is the dominant 
physical process determining the overall Z Cam standstill/outburst
pattern,
but that
viscous stochasticity provides an additional ingredient
explaining some of the secondary features observed.
\end{abstract}


\section{Introduction}

Dwarf novae (DNe), a subclass of cataclysmic variables, are binary
systems where matter overflows the Roche-lobe of a main sequence
star and forms a disk around a white dwarf primary.
Typically,
observations show large, regular variation in luminosity with
timescales between a few days and a few years, and with a magnitude
jump of between 2 and 6. It is believed that this variation issues
from a
thermal instability in the accretion disk owing
to profound
changes in the disk's opacity near the
ionisation temperature of hydrogen (H$\bar{o}$shi 1979, Meyer \& Meyer-Hofmeister
1981, Faulkner et al 1983, hereafter FLP). 

\begin{figure*}
\center
\includegraphics[width=12cm]{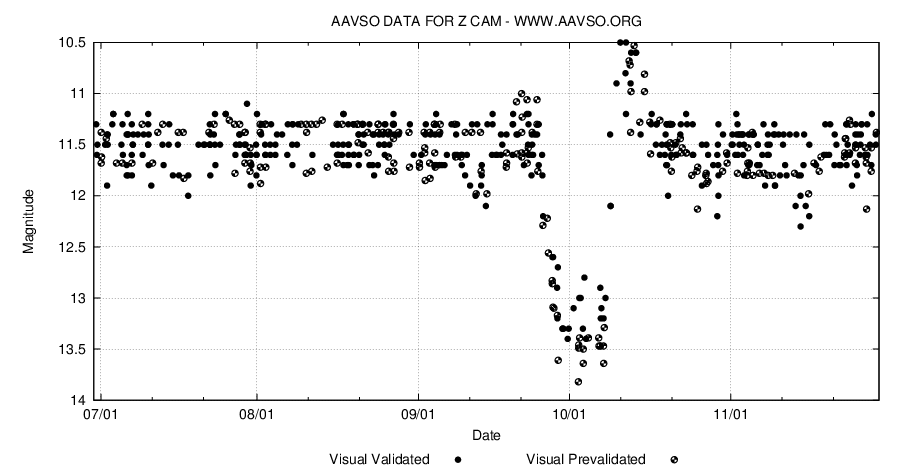}
\caption{Isolated outburst/dip in the lightcurve of Z Cam during a
  standstill (1981 July to 1981 December). Data from the AAVSO.}
\label{fig::ZCamLC}
\end{figure*}

One of the more intriguing sublasses of DNe are the Z Cam variables,
whose decay from outburst may be interrupted by long
\textit{standstills}. These correspond to periods where 
the luminosity remains roughly constant for
months to years (Warner 1995). Standstills are followed by a rapid drop to
quiescence and the initiation of a fresh
train of outbursts (Simonsen et
al, 2014). In some systems, standstills are punctuated by 
isolated `outburst/dip' pairs,
for example in the prototype itself, Z Camelopardalis. An example pair
is plotted in Figure
\ref{fig::ZCamLC}. The nova-like class of cataclysmic
variables also show occasional (though stunted) outbursts, 
such as UU Aqr, Q Cyg, and CP Lac
(Honeycutt et al 1998, Honeycutt 2001). 
Note that within the Z Cam group, there are outliers with even more
exotic behaviour. For example, the lightcurves of IW And and V513 Cas
show rises to outburst \emph{from} standstill (Simonsen et al.~2014).

It was understood early in the development of the disk instability model
that this behaviour could be explained by
variation in the mass transfer rate from the main sequence secondary
star (Meyer \&
Meyer-Hofmeister 1983, Lin et al. 1985, Buat-M\'{e}nard et
al. 2001, Lasota 2001). When the transfer is too great then the
system remains in the high state indefinitely; but if below a critical
value the system enters the recurrently outbursting state. Z Cam
variables are believed to straddle the bifurcation
point demarcating these two regimes, with variations in the transfer
rate pushing the system into one or the other. The critical rate
depends on parameters such as the binary mass ratio, disk viscosity,
disk radius, etc., and variations in the injection rate 
may be caused by irradiation of the
secondary (Smak, 1996), a solar type cycle (Livio \& Pringle,
1994), or starspots (King and Cannizzo 1998).

Turbulent stresses in the disk facilitate accretion of the transferred
material, while also providing the heating mechanism. 
But turbulence is fundamentally chaotic in
nature, producing fluctuations on the orbital timescale and longer
(e.g.\ Hawley et al.~1995), 
and these constitute an important feature of the flow. 
For example, the propagation of
such viscous fluctuations can explain the log-normality and
rms-flux relation in X-ray binaries and AGN (Uttley
\& McHardy 2001, Ingram \& van der Klis 2013, Cowperthwaite \&
Reynolds 2014). Another example concerns their effects on the thermal stability
of X-ray binaries (Janiuk \&  Misra 2012) and on thermal instability
in general (Ross et al 2017). In this paper we examine the role of
turbulent
fluctuations in triggering outbursts from standstill in DNe. For most mass transfer
rates, fluctuations do not endanger the coherence of the
limit cycle (see MRI simulations by
Latter \& Papaloizou 2012 and Hirose et al.~2014) 
and hence may be `smoothed out' with an eddy-viscosity model. However, for systems
perched on criticality, turbulent variations in the mass accretion
rate may easily instigate limit cycles, and can no longer be
neglected. 
Such noisy critical systems have multiple analogues, perhaps the
closest being the Fitzhugh-Nagumo (FHN) model for signal
transmission along nerves. Indeed, fluctuation-induced limit cycles  
have been observed and extensively studied in
that context (Treutlein \& Schulten 1985, 1986).

We employ two, relatively simple, 
models to explore this behaviour. 
The first is a one-zone representation of the surface density and
central temperature in a disk annulus at an outer radius. 
Essentially, it combines the surface density evolution equation of Mineshige \&
Osaki (1983) 
and the reduced thermodynamical approach of FLP, while adding
a stochastic alpha viscosity. The result is a noisy two-dimensional 
dynamical system, similar in most respects to the FHN equations. 
We also make use of a global one-dimensional viscous disk model, drawn
from the work of Papaloizou et al.~(1983), Lin et al.~(1985) and
Cannizzo (1993a), but which
includes a stochastically varying viscosity. It thus has some features
in common with the model of Janiuk \& Misra (2012), which describes
the radiation-pressure dominated flows around black holes.

Our results show that fluctuations alone can induce limit cycles from
standstills leading to individual, or occasionally repeated, 
outbursts as observed in some Z Cam
and nova-like cataclysmic variables. 
U Gem type behaviour is also produced for a mass transfer rate above
the classical critical value if the turbulent amplitude is
sufficiently large.
However, long trains of outbursts
separated by long standstills are
not a natural outcome of the simulations, 
unless we also include mass transfer variation
from the secondary. The results suggest that
turbulent fluctuations are an important, previously omitted,
ingredient in DN modelling, complementary to mass transfer
variability but by no means a replacement.

The plan of the paper is as follows. In Section 2 we introduce the 1D
viscous disk model for DNe. In Section 3 we reduce the problem to a 
one-zone model to illustrate the idea of stochastic excitability both with
and without mass transfer variability. Section 4 then demonstrates
that this local phenomenon translates to a quasi-global disk setting. 
We discuss our results in Section 5.  

\section{Viscous disk model}

One of the simplest ways of modelling an accretion disk is to
conduct a global vertical and azimuthal average. Then, with some
additional
weak assumptions, we obtain
a single
equation for surface density in radius and time. 
If the turbulence in the disk is modelled by an eddy viscosity,
the surface density
\begin{equation}
\Sigma(r)=\int^{\infty}_{-\infty}\rho dz,
\end{equation}
obeys the nonlinear diffusion equation (Pringle 1981),
\begin{equation}
\frac{\partial\Sigma}{\partial t}=
\frac{3}{r}\frac{\partial}{\partial r}\left[r^{1/2}
\frac{\partial}{\partial r}\big(r^{1/2}\nu\Sigma\big)\right] + S(r,t),
\label{Eqn::massDiff}
\end{equation}
where $\nu$ is the kinematic viscosity, in general a function of surface density,
midplane temperature $T_c$, and radius. The term $S$ represents the
source of mass from the secondary. The disk is assumed to be Keplerian.

In order to describe the evolution of the midplane temperature, 
we call on the
 vertically integrated, thin-disk,
energy equation 
\begin{equation}
C_{P}\frac{DT_{c}}{Dt}=\frac{9\nu\Omega^{2}}{4}-\frac{2F_{s}}{\Sigma}-
\frac{2H}{r\Sigma}\frac{\partial \left(rF_{r}\right)}{\partial r}
 -\frac{\mathcal{R}T_{c}}{\mu }\frac{1}{r}\frac{\partial (rv_{_{r}})}{\partial r}.
\label{eqn::Energy}
\end{equation}
We have used the form of this equation favoured by Cannizzo (1993a, see
also Lasota 2001).
The material derivative, $D(\cdot)/Dt$, includes the contribution from
radial advection, $v_{r}\partial(\cdot)/\partial r$, as well as the
local change with time, $\partial(\cdot)/\partial t$. The radial
advection speed is due to accretion and can be written in terms of the
viscous stress. The disk semi-thickness is $H$, $C_P$ is the
specific heat, $\mu$ is molecular mass, and $\mathcal{R}$ is the gas
constant. 

The first term
on the right hand-side is the local vertically integrated viscous
dissipation. This is followed by the local radiative cooling rate, 
where $F_{s}$ is the flux emitted by each disk
surface. The next describes radial radiative transport of energy, 
where $F_{r}$ is the radiative flux in the radial direction, given by
\begin{equation}
F_{r}=-\frac{2H\sigma T^{3}}{\kappa\Sigma}\frac{\partial T_{c}}{\partial r},
\label{eqn::radFlux}
\end{equation}
where $\kappa$ is the opacity. 
The last term on the right hand side represents
 the energy released by $PdV$ work where $\mathcal{R}$ is the ideal gas constant.

In order to calculate the radiative cooling at the surfaces, 
the disk's vertical structure must be determined at each radius (Lasota, 2001),
necessitating the inclusion of radiative transport, the gas's
ionisation state, etc. Our aim, however, is
to sketch out the qualitative behaviour of the dynamics rather than
match observations in detail, and so we adopt
the simpler, somewhat more
transparent approach, of
FLP who model the dependence of cooling rate on 
$T_c$ via a set of analytical, physically motivated prescriptions which
we now summarise.

It is assumed that the disk is emitting as a black body and so $F_{s}$ is given by
\begin{equation}\label{Fs}
F_{s}=\sigma T_{e}^{4},
\end{equation}
where $\sigma$ is the Stefan-Boltzmann constant and $T_{e}$ is the
effective temperature, related to $T_{c}$ by taking into account
the vertical structure of the disk. The connection between $T_{e}$ and
$T_{c}$ may be determined by
considering three temperature regimes, distinguished by the ionization 
 fraction of hydrogen. For each regime a different parametrisation for $T_{e}$ is necessary. 
\begin{enumerate}
\item The first is when the disk is mostly ionised and is in the hot optically thick
  state, with $T_c>10^4$ K. 
 In this case the majority of the optical depth originates from near
  the mid-plane. 
 This leads to the relation
\begin{align}
T_{e1}^{4}=\frac{4}{3}\frac{T_{c}^{4}}{\tau_{c}},
\end{align}
where $\tau_{c}$ is the mid-plane optical depth given by
$\tau_{c}=\kappa_{c1}\Sigma$.
 The opacity in regime 1 may be approximated by Kramer's formula,
\begin{align}
\kappa_{c1}=1.5 \times 10^{20}\frac{\Sigma}{2H}T^{-2.5}.
\label{eq::kapa1}
\end{align}
\item The next regime is when hydrogen is partially ionized in the
  mid-plane but the cold photosphere is mostly neutral. There is a
  sharp drop-off in ionisation with height when the temperature drops
  though  the ionisation threshold.
  An abrupt change in opacity results 
  and the dependence of effective temperature $T_{e2}$ on central
  temperature
  $T_{c2}$ becomes very weak. In fact, we have
\begin{align}
T_{e2}=\left(10^{36}E\left(\frac{\Sigma}{2H}\right)^{-1/3}/\Sigma\right)^{1/10},
\end{align}
where $E$ is dimensionless constant that 
can be optimised to improve the approximation. 
\item The last regime is when the whole vertical extent of the disk is
  cold and poorly ionised, $T_c< 4000$ K. Then we may approximate the
  relationship between effective and central temperatures by
\begin{align}
T_{e3}=(2\lambda\tau_{c})^{1/4}T_{c},
\end{align}
where $\lambda$ is another dimensionless constant. As in regime (ii), the
opacity is dominated by H$^-$ ions and may be approximated by
\begin{align}
\kappa_{c3}=10^{-36}\left(\frac{\Sigma}{2H}\right)^{1/3}T^{10}.
\label{eq::kapa3}
\end{align}
\end{enumerate}
In the above $3$ expressions for $T_{ci}$, the quantities are 
in CGS units. To prevent discontinuities when matching these three expressions, we adopt 
\begin{equation} 
T_{e}^{4}=\frac{\left(T_{e1}^{4}+T_{e2}^{4}\right)T_{e3}^{4}}{\left(\sqrt{\left(T_{e1}^{4}+T_{e2}^{4}\right)}+T_{e3}^{2}\right)^{2}}.
\label{Eqn::match}
\end{equation}
For a more detailed discussion of the radiative cooling model we
 refer the reader to FLP and Latter \& Papaloizou (2012).
We choose $E=5.66$ and $\lambda=0.5$, as in FLP.

Note that the FLP model is a crude simplification of the full vertical
radiative transfer problem, and performs poorly on the lower stable
branch of solutions. However, on the hot upper branch, in
which we are primarily interested in this paper, it is more reliable.  
Some of the model's peculiarities include (a) an exaggerated `elongation'
of the S-curve in the $\Sigma$ axis: the two corners of the `S' differ
by a factor of 3 or more, in contrast to more realistic curves obtained from
vertical integrations (e.g.\ Meyer \& Meyer-Hofmeister 1981);
 (b) the scalings of these corners with disk
parameters, such as $\alpha$, also differ from the vertical
integrations (e.g.\ Hameury et al.~1998); (c) detailed
features of obtained light curves are not reproduced correctly.
See Cannizzo (1993b) for criticisms of the model. That all being said,
for our purposes (proof of concept) the FLP
approach is more than adequate, and furthermore very convenient.

To close the system of equations, we adopt 
the $\alpha$-model to express $\nu$ in terms of the other variables,
and so explicitly link the two evolution equations  (Shakura \&
Sunyaev, 1973). We stipulate that
\begin{equation}
\nu=\frac{2}{3}\alpha\frac{\mathcal{R}T_{c}}{\mu\Omega}.
\label{eqn::alpha}
\end{equation}
 In this work $\alpha$ is a
 stochastic 
function of time through which we incorporate the effects of
fluctuations in stress. The details of how this is modelled are given
in the following sections.

In summary, the partial differential equations
\eqref{Eqn::massDiff}-\eqref{eqn::Energy}, 
along with the supplementary equations \eqref{eqn::radFlux} - 
\eqref{eqn::alpha}
and appropriate boundary conditions, approximate the time evolution of
a 1D dwarf nova disk.


\section{Local one zone model}

\begin{figure}
\centering
\includegraphics[width=9cm]{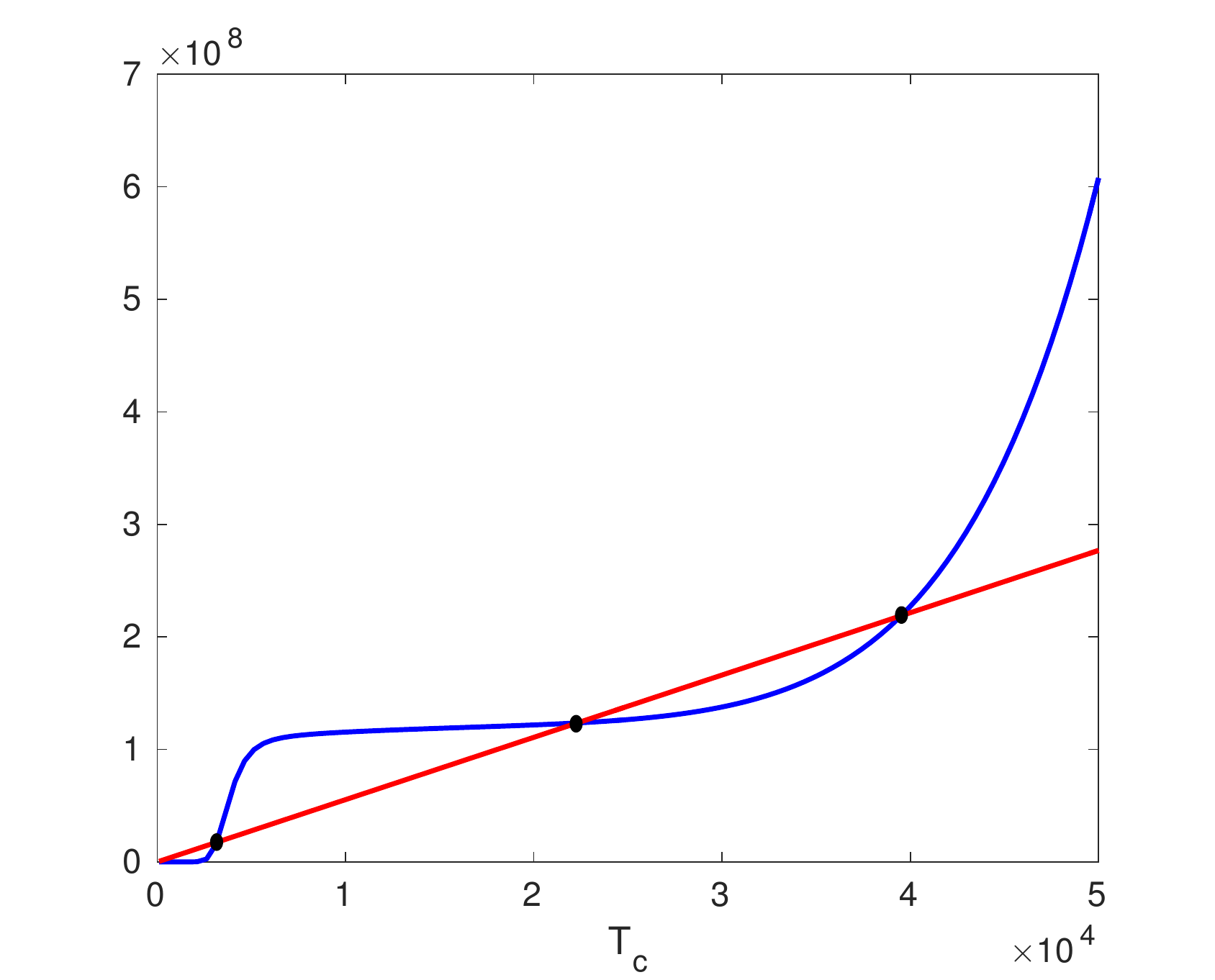}
\caption{Cooling, $\Lambda$ (blue) and heating, $\mathcal{H}$, (red)
  as functions of $T_{c}$ for $\Sigma=300\text{g cm}^{-2}$ and
  $\Omega=2.22\times10^{-3}s^{-1}$. The black dots correspond to the
  $3$ thermal equilibria.}
\label{fig::heatcool}
\end{figure}  

\begin{figure}
\centering
\includegraphics[width=9cm]{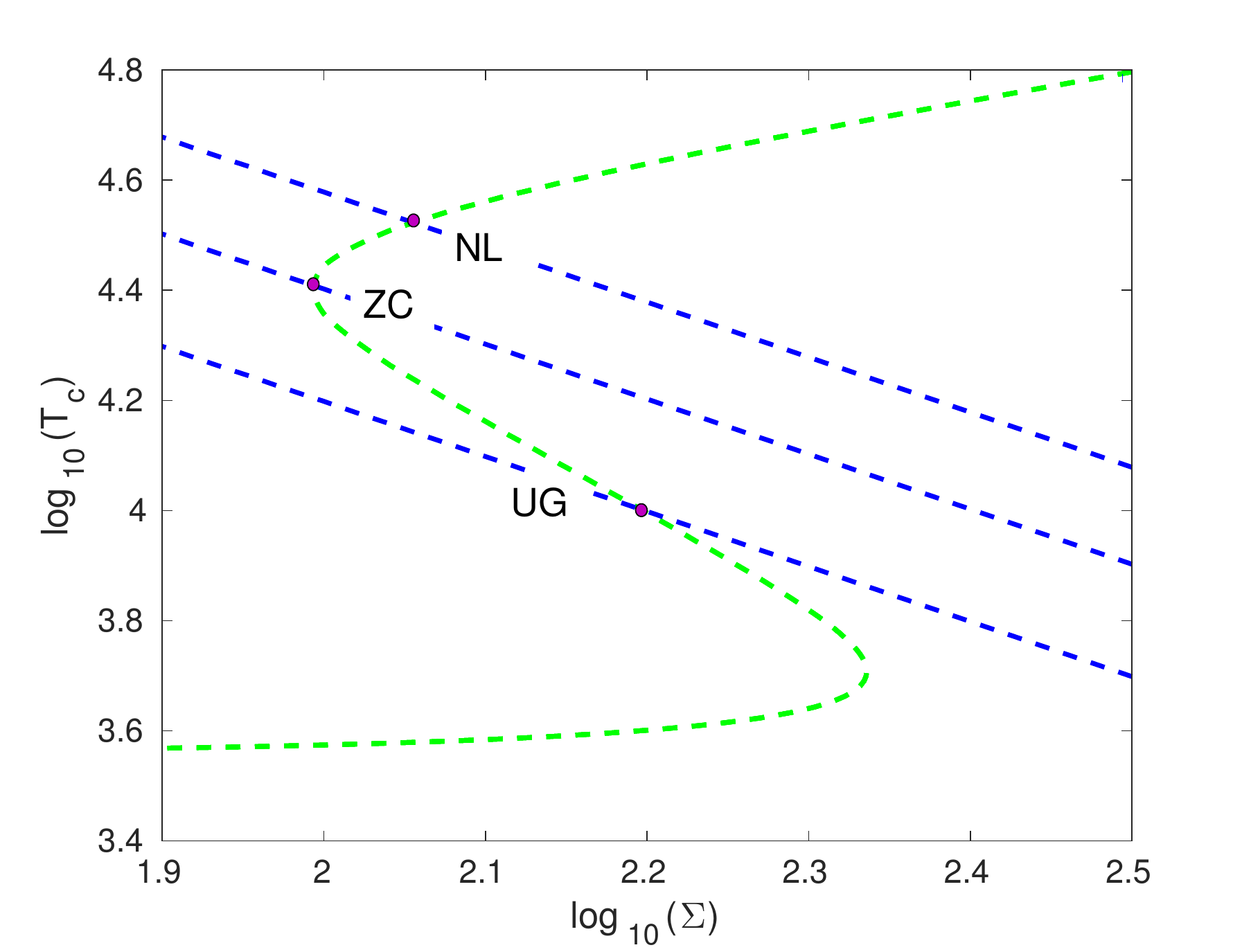}
\caption{A single $T$-nullcline (green) and three $\Sigma$-nullclines (blue)
  corresponding to three different mass injection rates. Intersections
of the curves correspond to fixed points of the system. Fixed points
on the upper branch are stable, and the disk is always in the hot,
highly accreting state. Such systems may correspond to Nova-like
variables (NL). Fixed points on the middle branch are unstable and the
system undergoes limit cycles. Such systems correspond to U Gem
variables (UG). If the intersection point is just above the left corner of
the $T$-nullcline (the saddle node), then perturbations to the accretion rate can easily move the
$\Sigma$ nullclines to the unstable middle branch precipitating a
train of outbursts. Such systems are identified with the Z Cam class
of variables (ZC). The three mass injection rates are 6, 4, and 2.5 $\times
10^{16}$ g s$^{-1}$ respectively.}
\label{fig::Scurves_radius}
\end{figure}

\subsection{Reduced dynamical system}

\subsubsection{Evolutionary equations}

To fix ideas 
we begin our discussion with a 
reduced system. Consider an annulus at fixed radius
$r=R$ and radial extent $\Delta R$. Next assume that
mass is injected into the annulus at rate $\dot{M}$ and that its
neighbouring annuli are in steady state. 
Mineshige and Osaki (1983) derive a convenient first order 
ODE for the surface mass density of such an annulus
\begin{equation}
\frac{d \Sigma}{d t} = -2\left(\frac{R}{\Delta R}\right)^{2}\frac{W-W_{0}}{(GMR)^{1/2}},
\label{eqn:mass}
\end{equation}
where $M$ is the mass of the central star and
$W=2\alpha P H$ is the vertically integrated viscous stress.
We denote by $W_{0}$ its value in steady state,
\begin{equation}
W_{0}=\frac{(GM)^{1/2}}{2\pi R^{3/2}}\left[1-\left(\frac{R_{*}}{R}\right)^{1/2}\right]\dot{M},
\end{equation}
which is the stress the disk requires in order to process the injected mass
from the secondary without any local build up (or evacuation) of mass. 
(Here $R_{*}$ is the radius of the white dwarf.) If the actual
stress in the annulus $W$ is below or greater than $W_0$ then the mass
of the annulus changes.

Using the $\alpha$-prescription, Equation \eqref{eqn:mass} can be
rewritten as the nonlinear ODE 
\begin{equation}
\frac{d \Sigma}{d t} = a_1 - a_2\alpha \Sigma\,T_c,
\label{eq:0Ddif}
\end{equation}
where the constants $a_1$ and $a_2$ are given by 
\begin{align}
a_1=\frac{\dot{M}}{\pi(\Delta
    R)^2}\left[1-\left(\frac{R_{*}}{R}\right)^{1/2}\right], \quad
  a_2= \frac{\mathcal{R}}{\mu\Omega(\Delta R)^{2}},
\end{align}
and $\alpha$ may depend on time.

Turning now to the energy equation, we neglect advection,
heating through compression, and radial radiative transport. 
The change
in temperature of the annulus is determined solely from the competition
between the viscous heating, $\mathcal{H}$, and radiative cooling,
$\Lambda$, defined below. The thermal energy equation \eqref{eqn::Energy} is then 
\begin{align}\label{eq:0DTherm}
C_{P}\frac{d T_{c}}{d t}&= a_3\alpha\, T_c -\frac{2}{\Sigma} F_s(T_c,\,\Sigma),
\\ &=\mathcal{H}-\Lambda,
\end{align}
where the constant $a_3$ is
\begin{align}
a_3 = \frac{3}{2}\frac{\Omega\mathcal{R}}{\mu}.
\end{align}
The surface flux of energy $F_s$ is computed from 
Eqs \eqref{Fs}-\eqref{Eqn::match}.

The viscous variation of $\Sigma$ occurs on a timescale of 
\begin{equation}
t_{\Sigma}=\frac{(\Delta R)^{2}}{\alpha H^{2}\Omega},
\end{equation}
while the thermal timescale is roughly $t_\text{th}\sim
1/(\alpha\Omega)$, and is hence $\varepsilon^2=H^2/(\Delta R)^2 \ll 1$ shorter.
The dynamical system then exhibits `fast-slow' dynamics and may be
described by dimensionless equations
\begin{align*}
\frac{d\Sigma}{dt}= f(\Sigma,\,T_c,\,t), 
\qquad \frac{dT_c}{dt} = \frac{1}{\varepsilon^2}g(\Sigma,\,T_c,\,t),
\end{align*}
for nonlinear functions $f$ and $g$, derived from Eqs \eqref{eq:0Ddif} and
\eqref{eq:0DTherm}. 
The second (thermal) equation varies much
faster than the first (density) equation.
While the surface density evolves on $t_{\Sigma}$,
the system remains near to thermal equilibrium, 
hugging the curve given by $g\approx 0$ in the phase
space of $(\Sigma,\,T_c)$. Only when thermal equilibrium is lost does
the second equation `activate' and there is rapid variation on
$t_\text{th}$: an outburst or drop to quiescence.

\subsubsection{Nullclines}

In Figure \ref{fig::heatcool} we plot examples of heating and cooling
curves for an annulus in the outer disk at fixed $\Sigma$ and constant
$\alpha$. The
non-zero intersections give the thermal equilibria, $d
T_{c}/d t=0$, of which there can be $1,2$ or $3$ for a given
$\Sigma$. 
Likewise, there exists a
surface density equilibrium solution, $d\Sigma/d t=0$, which, at a
given $\Sigma$, only ever has one solution. When
considering the $\Sigma-T_{c}$ phase-space, the thermal equilibrium
curve is $S$-shaped while the surface density equilibrium curve is a
monotonically decreasing function of $\Sigma$.  For brevity we will
henceforth refer to the surface density equilibrium solution as the
$\Sigma$-nullcline and the thermal equilibrium curve as the
$T$-nullcline. 

The intercept of the two nullclines corresponds to a true
fixed point of the dynamical system.  
The disk is capable of processing the mass it
receives in a steady manner, and energy is produced and leaves the
system
at the same rate.
If the intersection occurs on the upper (or lower) branch of
the S-curve then the equilibrium is stable and the system remains
there indefinitely. If the intersection occurs on the middle
branch then the equilibrium is unstable, and if slightly perturbed
the system falls into a
pattern of periodic orbits (limit cycles). Such systems are
identified with the classical U Gem class of variables. 
There are two bifurcation
points, located at the two corners of the S-curve (the saddle
nodes). As a control parameter is varied, $\dot{M}$ or possibly $R$,
the intersection point may drop from the upper branch, pass through
the left-most saddle point, and then fall upon the unstable middle
branch. Such a transition corresponds to a `global Hopf bifurcation',
and the system moves from a steady state to an outbursting state:
the kind of 
dynamics we associate with a Z Cam variable. In Figure 
\ref{fig::Scurves_radius} we plot $\Sigma$ and $T$ nullclines for
various $\dot{M}$, giving fixed point that are stable, marginal, and
unstable. 

\subsubsection{Stochasticity}

The model permits the inclusion of time-variability in either or both
the mass transfer rate and the turbulent viscosity.
A low amplitude time-dependent mass transfer may be modelled by
a square wave in $\dot{M}$, oscillating around $\dot{M}_0$ 
with amplitude $\delta\dot{M}_0$ and period $t_I$. Here $\delta$ is a
dimensionless parameter.  
Note in this case, the derivation of Equation
\eqref{eqn:mass} requires the disk to be in a steady state, and so
$W_0$ must be redefined replacing $\dot{M}$ with $\dot{M_0}$.

\begin{figure*}
\includegraphics[width=8cm]{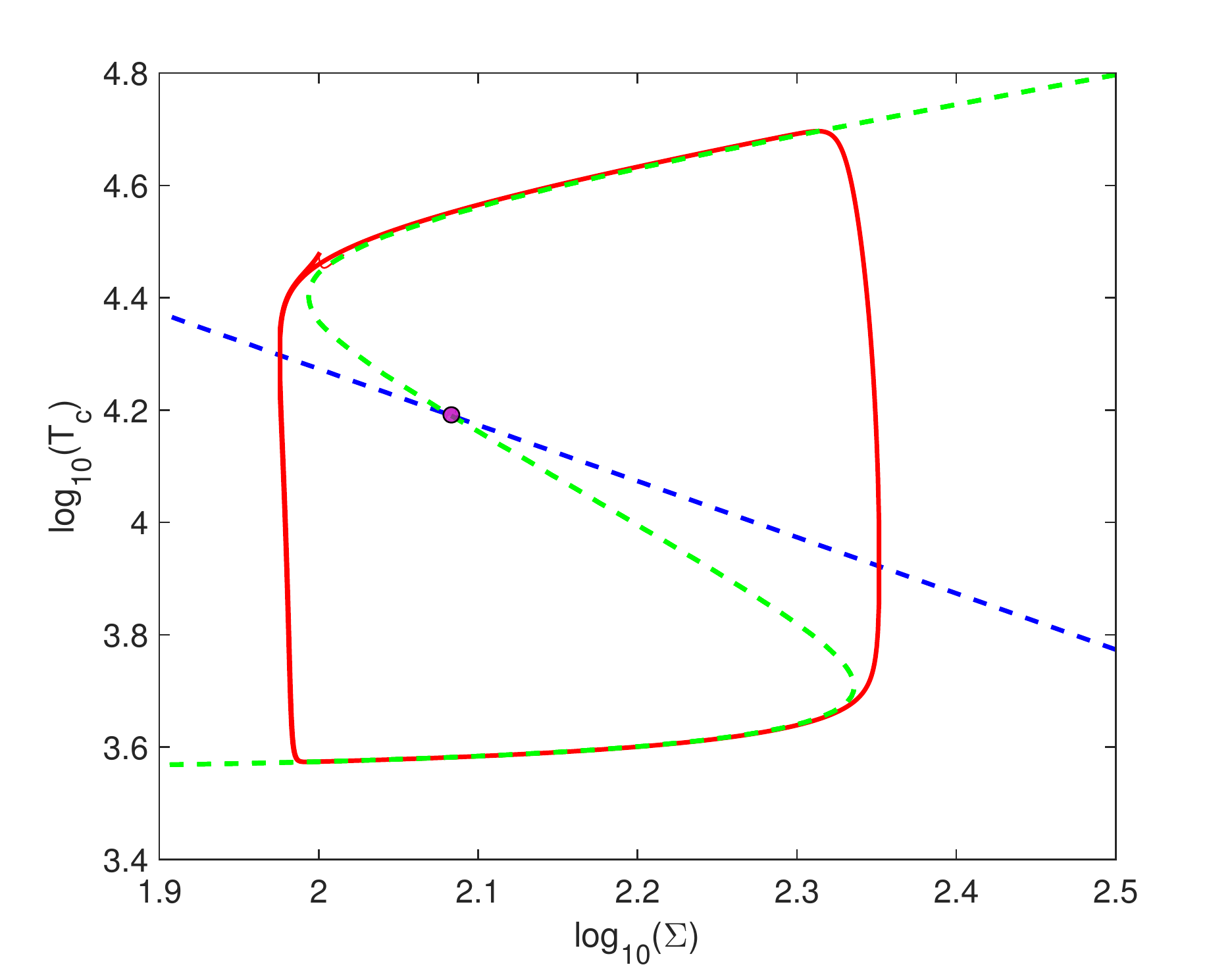}
\includegraphics[width=8cm]{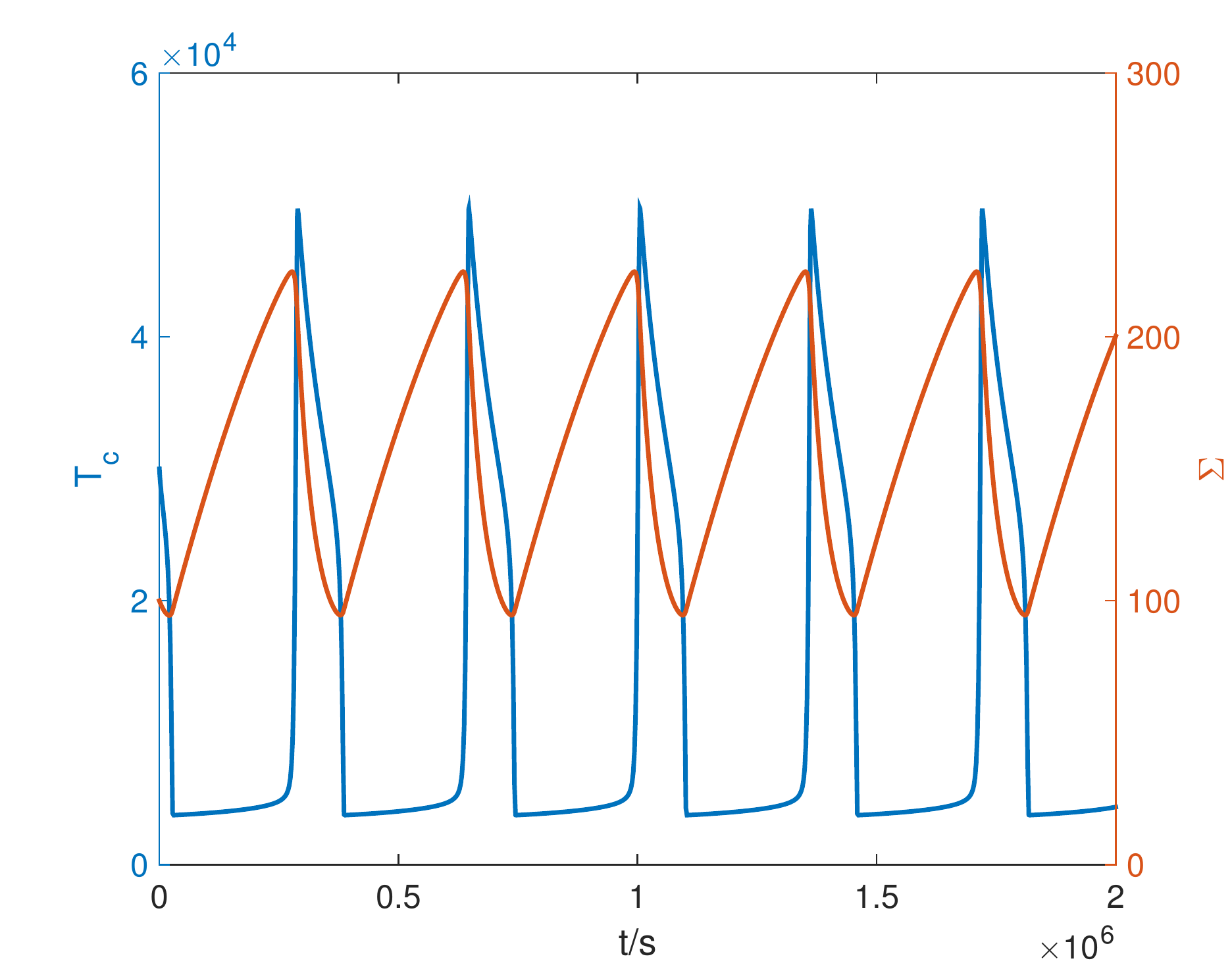}
\caption{Both panels correspond to a `U Gem simulation' with
  $\dot{M}=3.8\times10^{16}\text{g s}^{-1}$ and $b_{0}=0$. 
Left panel (a): Trajectory of a disk annuli through the $\Sigma-T_{c}$
phase space (red).
The thermal equilibrium curve (dashed green curve) and mass nullcline 
(dashed blue line) are superimposed. The fixed point is shown by the purple dot.
Right panel (b): Evolution of $T_{c}$ and $\Sigma$ as functions of time. }
\label{fig::UGem}
\end{figure*}

To account for viscous fluctuations 
we introduce a time-varying term to $\alpha$ (Lyubarskii 1997) 
\begin{equation}
\alpha\left(t\right)=\alpha_{0}\left[1+\beta\left(t \right)\right].
\label{Eqn::0Dfluctuations}
\end{equation}   
In this equation $\alpha_{0}$ is a 
constant value assumed to be $0.1$ while $\beta(t)$, after time
discretization, 
is the fluctuating term given by the Markov chain
\begin{equation}
\beta^{n}=b_{0}u^{n},
\label{Eqn::0Dbeta}
\end{equation}
where $b_{0}$ is a parameter that determines the noise amplitude. The subscript $n$ gives the time-step in fluctuation times and
\begin{equation}
u^{n}=\tilde{\alpha}u^{n-1}+\epsilon^{n},
\label{Eqn::0Dun}
\end{equation}
where $\tilde{\alpha}$ is another parameter chosen with the requirement
that $\left|\tilde{\alpha}\right|<1$, to ensure that the Markov chain does
not diverge, and $\epsilon^{n}$ is a random variable taken from a uniform
distribution on the range $\left(0,1\right)$. This process has been
used in previous studies to model disk fluctuations (King et al.~2004,
Janiuk \& Misra 2012), but other simpler prescriptions would also have
sufficed. 
The mean number of timesteps between oscillation peaks is given by 
\begin{equation}
N_{\text{step}}=\frac{2\pi}{\cos^{-1}\left[-1\left(1+\tilde{\alpha}\right)/2\right]}
\end{equation}   
and the amplitude of the oscillations is such that 
\begin{equation}
\text{Var}\left(u^{n}\right)=\frac{\text{Var}\left(\epsilon^{n}\right)}{1-\tilde{\alpha}^{2}},
\end{equation} 
see King et al.\ (2004) for more details. 
We use $\tilde{\alpha}=-0.5$ which gives $N_{\text{step}}\approx2.6$ and
$\text{Var}(u^{n})=0.25$. 

For illustrative purposes, later in this section we also employ a white
noise model for $\beta$, so that $\beta=b_0dW$, where $dW$ is a Wiener process.

\begin{figure*}
\includegraphics[width=16cm]{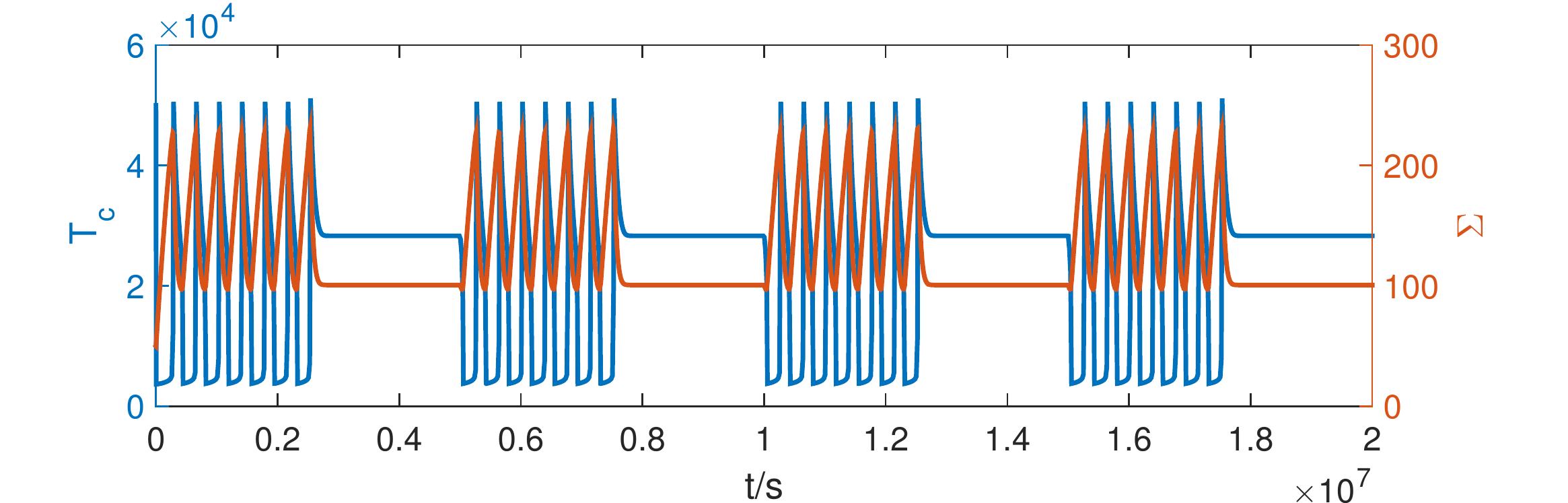}
\caption{Time evolution of the surface density and central temperature
  with $\dot{M}_0=4\times10^{17}\text{g s}^{-1}$ and $b_{0}=0$ with mass
  transfer variation. The variation amplitude is $\delta=0.1$ and
  $t_I=5\times 10^6 s$. The system clearly switches between outburst
  trains and standstills.}
\label{fig::ODZCam}
\end{figure*}

\subsection{Results: classical}

Before we approach the stochastic case, we consider the deterministic
non-fluctuating system ($\beta=0$) to  illustrate thermal limit cycles representative
of U Gem variables and, if mass transfer is included, Z Cam
variables. 
In this section we set the central mass $M=M_{\odot}$, 
the annulus width $\Delta R=3\times10^{9}\text{cm}$, 
$R=2.8\times 10^{10}\text{cm}$, and $R_{*}=3\times10^{9}\text{cm}$.

\begin{figure}
\includegraphics[width=8cm]{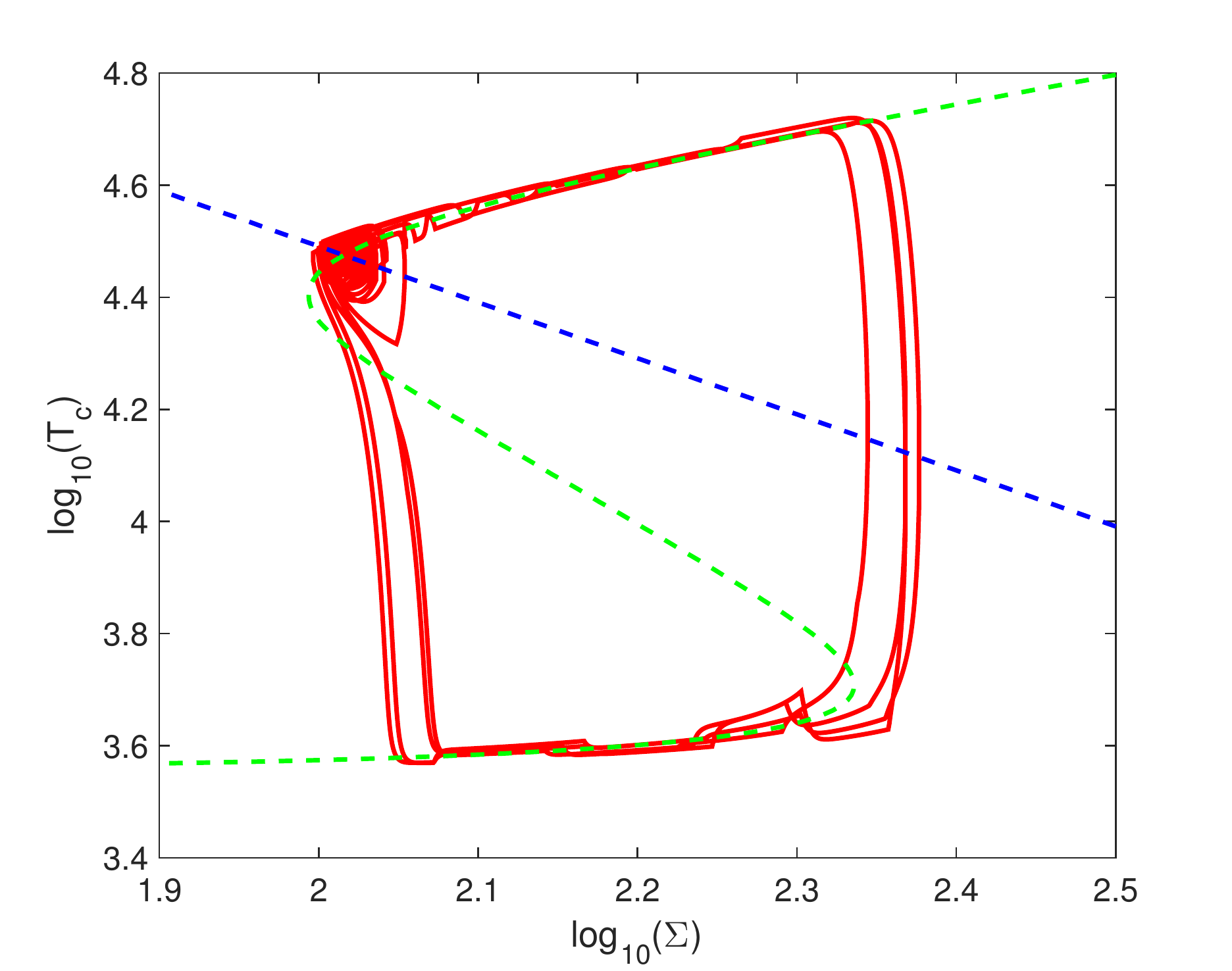}
\caption{Time evolution (red) of a disk annuli through the
  $\Sigma-T_{c}$ 
phase space. For the simulation, $\dot{M}=5\times10^{16}\,\text{g s}^{-1}$ and $b_{0}=0.1$.}
\label{fig::eg}
\end{figure}

\subsubsection{U Gem type simulations}

In a U Gem system, $\dot{M}$ must be sufficiently small such that the
$\Sigma$-nullclines, i.e. steady solutions to Equation \eqref{eq:0Ddif}, intersect
the $T$-nullcline on the middle unstable branch at some disk
radius. Let us centre our annulus at such a radius. Figure
\ref{fig::UGem}a exhibits a phase plane with this property. 
Here the $T$ and $\Sigma$ nullclines are given by the green and blue
curves, respectively.
The purple fixed point, their intersection, is unstable
and so the system admits periodic orbits. 

Let us assume that the disk annulus
is initially on the low branch. Mass is being fed into the annulus at
a greater rate than it is able to process it and so mass
accumulates and $\Sigma$ increases. 
Eventually, the annulus reaches the low temperature
bifurcation point and leaps to the high hot state. The
annulus, now above the $\Sigma$-nullcline, begins losing mass
because its accretion rate exceeds the supply rate. Once the high
temperature bifurcation point is reached, the annulus drops to the
low temperature state and the process is repeated. 

The red curve in
Fig.~\ref{fig::UGem}a 
corresponds to a numerical simulation of Eqs \eqref{eq:0Ddif} 
and \eqref{eq:0DTherm} and
illustrates this basic behaviour quite clearly.
In Figure \ref{fig::UGem}b we plot the evolution of $\Sigma$ and $T_c$ 
as a function of time. Here one can observe the two time-scales at
work, especially in the (blue) $T_c$ curve. In between outbursts there
is litle change in temperature, but a build up in mass (seen in the red curve)
occuring on the slow viscous timescale $t_{\Sigma}$. Interleaving these phases are
the abrupt bursts occuring on the short thermal timescale $t_{\text{th}}$.

\subsubsection{Z Cam type simulations}

Z Cam variables intermittently behave in a similar manner to U Gem
variables but occasionally remain on the high temperature branch
in a quasi-steady state for months to years before dropping to
quiescence and undergoing a fresh batch of outbursts. 
This is attributed to a shift in the $\Sigma$-nullcline
upwards (due to a change in mass transfer, $\delta\neq 0$),
so that the intersection is beyond the saddle node. Of course, the
system must be close to criticality if this is to be possible.

To demonstrate this in our model, $\dot{M_{0}}$ and $\delta$ are 
chosen such that the
fixed point oscillates between the stable high temperature
branch and the unstable branch. When the fixed point is on the upper
branch the system is stable and will remain in a standstill. A
decrease in $\dot{M_{0}}$ pushes the fixed point onto the unstable
branch and the system undergoes periodic orbits resulting in trains of
outbursts. In Figure \ref{fig::ODZCam} we  
show an example of standstill-outburst-train cycles.

\begin{figure*}
\includegraphics[width=8cm]{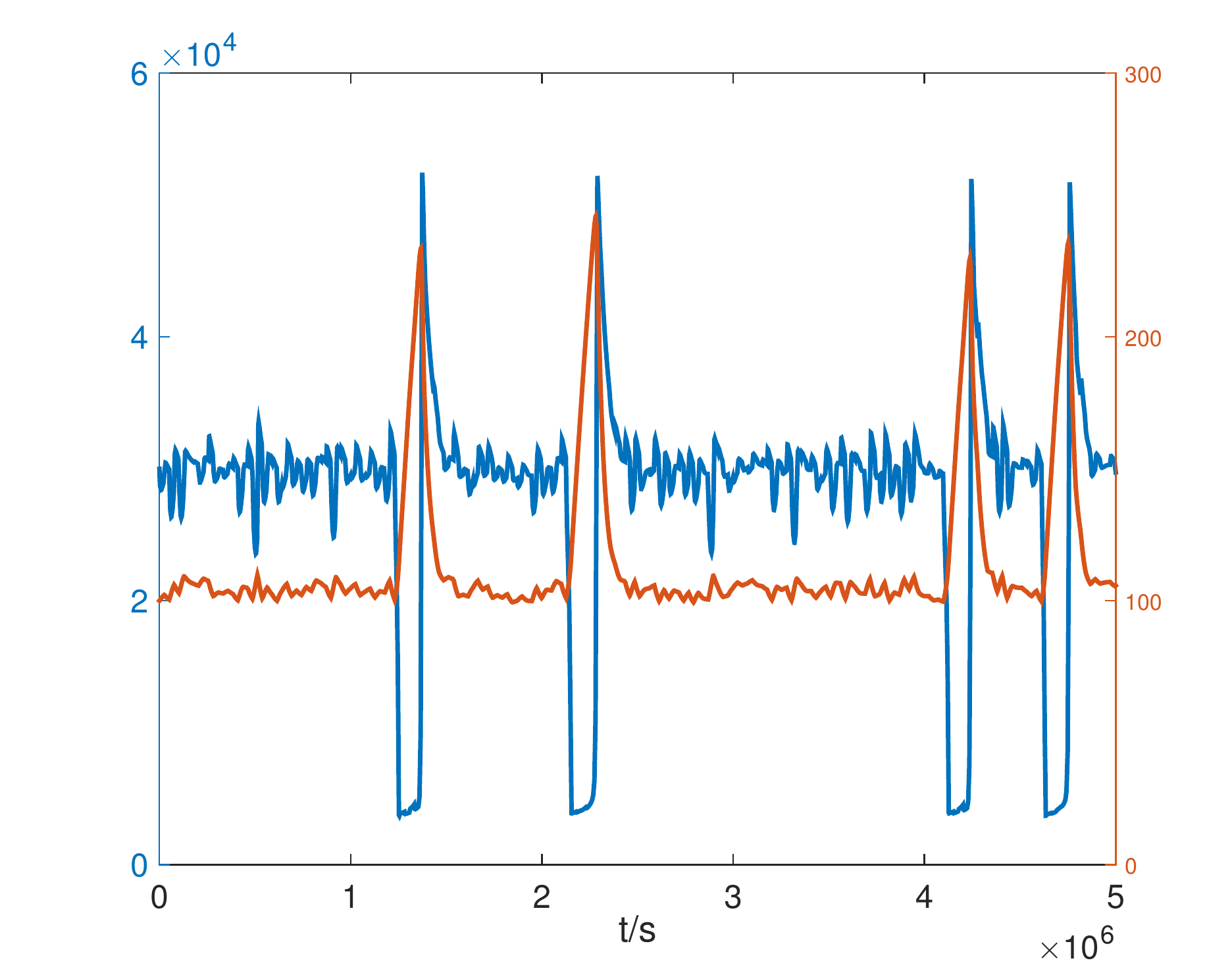}
\includegraphics[width=8cm]{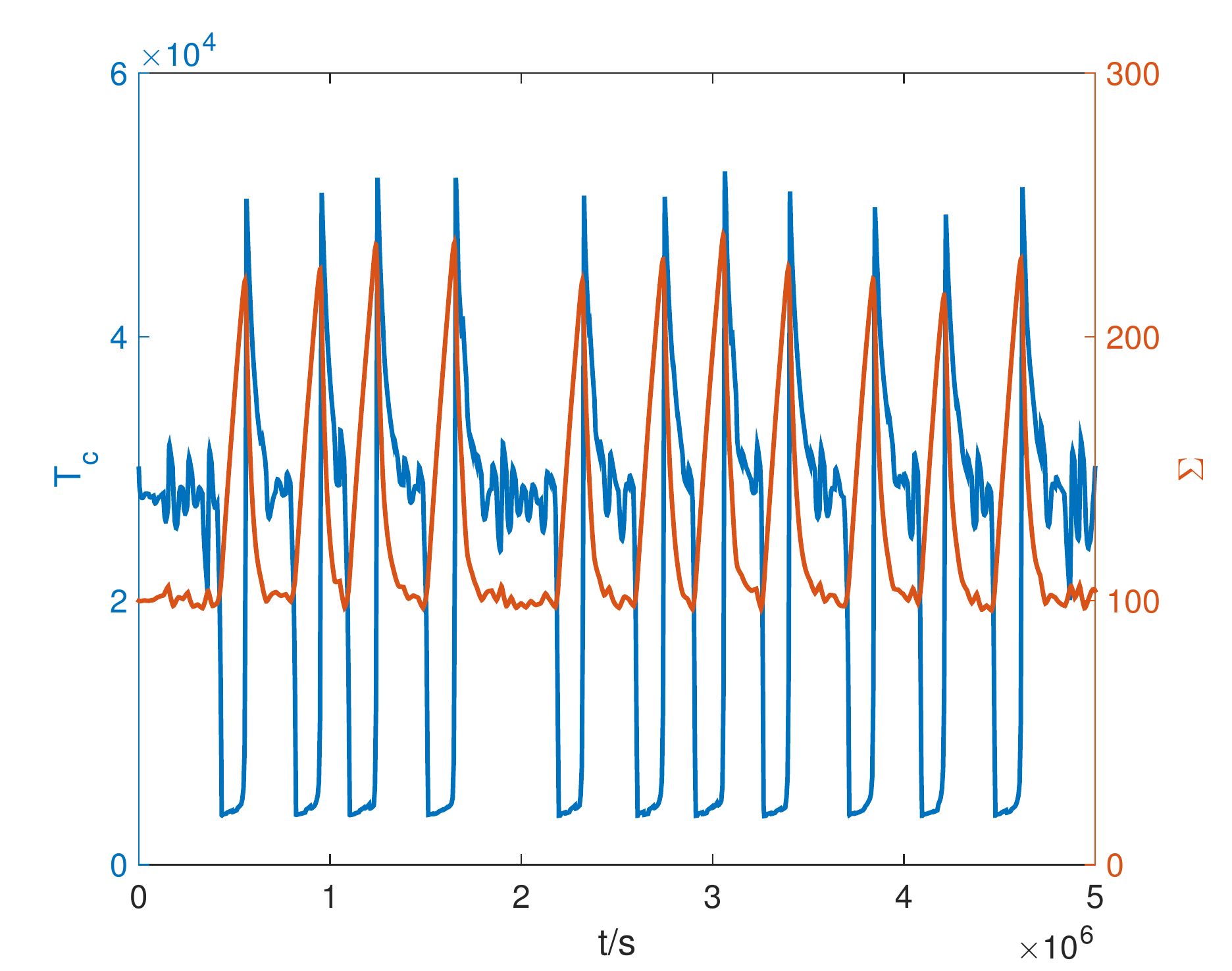}
\includegraphics[width=8cm]{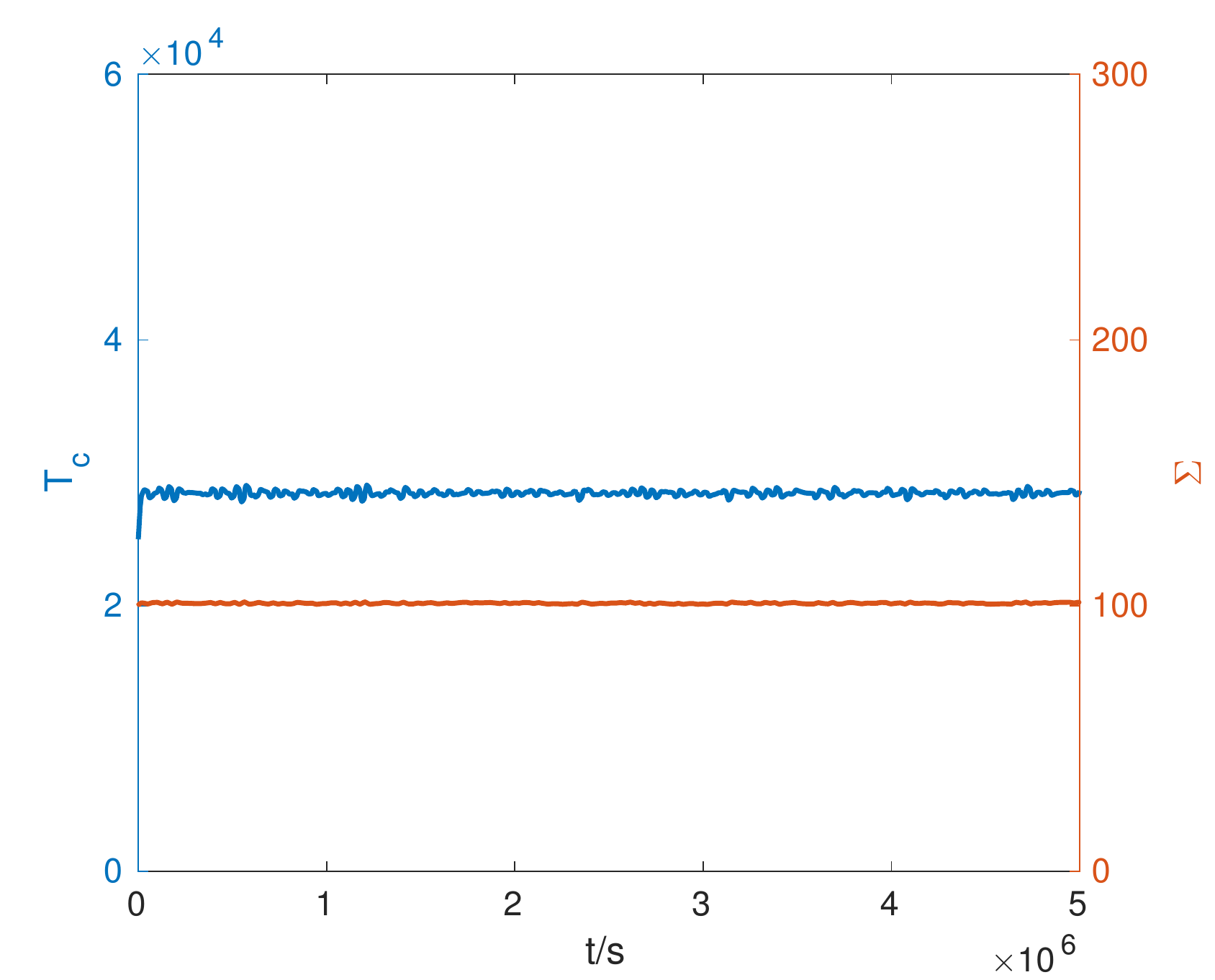}
\includegraphics[width=8cm]{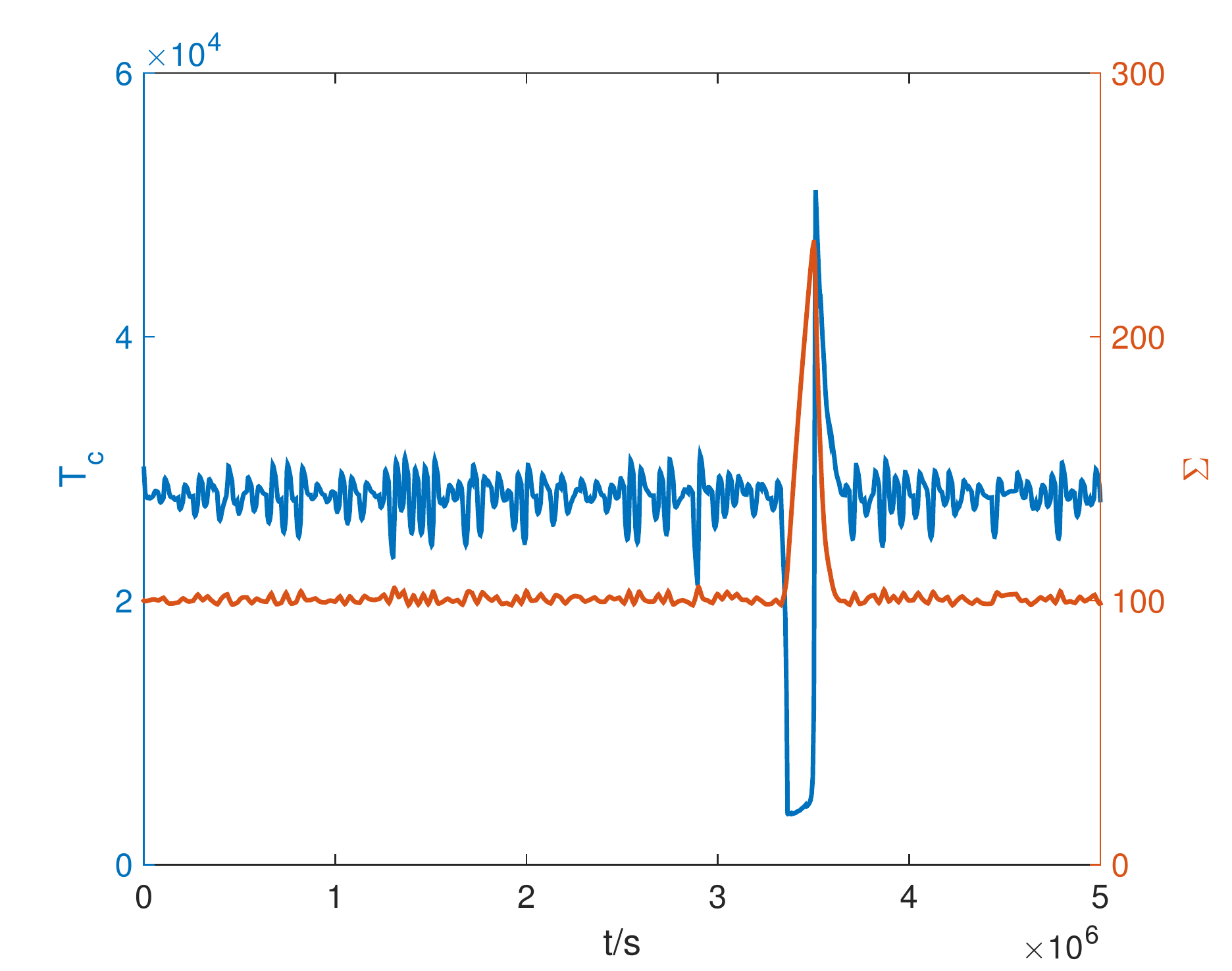}
\caption{ Evolution of central temperature $T_c$ (blue) and surface
  density $\Sigma$ (red) as functions of time. 
 Top left panel (a) has $\dot{M}=5\times10^{16}\,\text{g s}^{-1}$ and top right
 panel (b) $\dot{M}=4.55\times10^{16}\,\text{g s}^{-1}$, 
both with $b_{0}=0.1$.
Bottom panels have $\dot{M}=4.55\times10^{16}\,\text{g s}^{-1}$ with (c)
$b_{0}=0.01$ (left) and (d) $b_{0}=0.05$ (right). }
\label{fig::0DMdot}
\end{figure*}

\subsection{Results: stochastic}

With the basics established, we include a fluctuating viscosity
($\beta\neq0$). The fluctuation time scale is taken to be
$10t_{orb}$, which is motivated  
by MRI shearing box simulations (Hawley et al 1995). 
The amplitude is set at $b_{0}=0.1$.
First, we consider the case where there is no mass
transfer variability ($\delta=0$). To obtain a fixed point on the
high temperature branch close to the bifurcation point we choose
$\dot{M}=5\times10^{16}\text{g s}^{-1}$. A deterministic system with this
configuration would remain in the high temperature state
indefinitely. A stochastic system, however, will see the fixed point
jostle about as the two nullclines fluctuate with $\alpha$, sometimes
falling on the upper branch, sometimes dropping to the middle branch
of the S-curve. The
$\Sigma$-nullcline is perhaps most sensitive to viscous fluctuations because
equilibria upon it obey $T_c\sim\alpha^{-1}$.

A series of 4 stochastically excited outbursts is plotted 
in the phase space of $\Sigma$--$T_c$ in Figure \ref{fig::eg}, and
as a function of time in Figure \ref{fig::0DMdot}a. 
The disk annulus initially hangs around the fixed point for
$\approx 1.2\times10^{6}s$ before being stochastically excited,
transitioning to the cooler branch, and undergoing limit
cycles. During the dynamical phase of the cycle, there are small
fluctuations around the stable manifold due to the variable $\alpha$
but they have little effect on the overall dynamics. It is when the
system is close to the fixed point that the fluctuations become
important, ultimately causing excitation of the system.

Note the slightly delayed
rise to outburst at the bottom right corner of the S curve in 
Figure \ref{fig::eg}, near the saddle node. This is a delay brought
about by the stochastic noise, and was witnessed in earlier MRI
simulations (Latter \& Papaloizou 2012, Hirose et al.~2014). The
system, of course, is held up for much longer at the top left corner,
because of the stable fixed point. This region is discussed in more
detail in the next subsection.

\begin{figure}
\includegraphics[width=8cm]{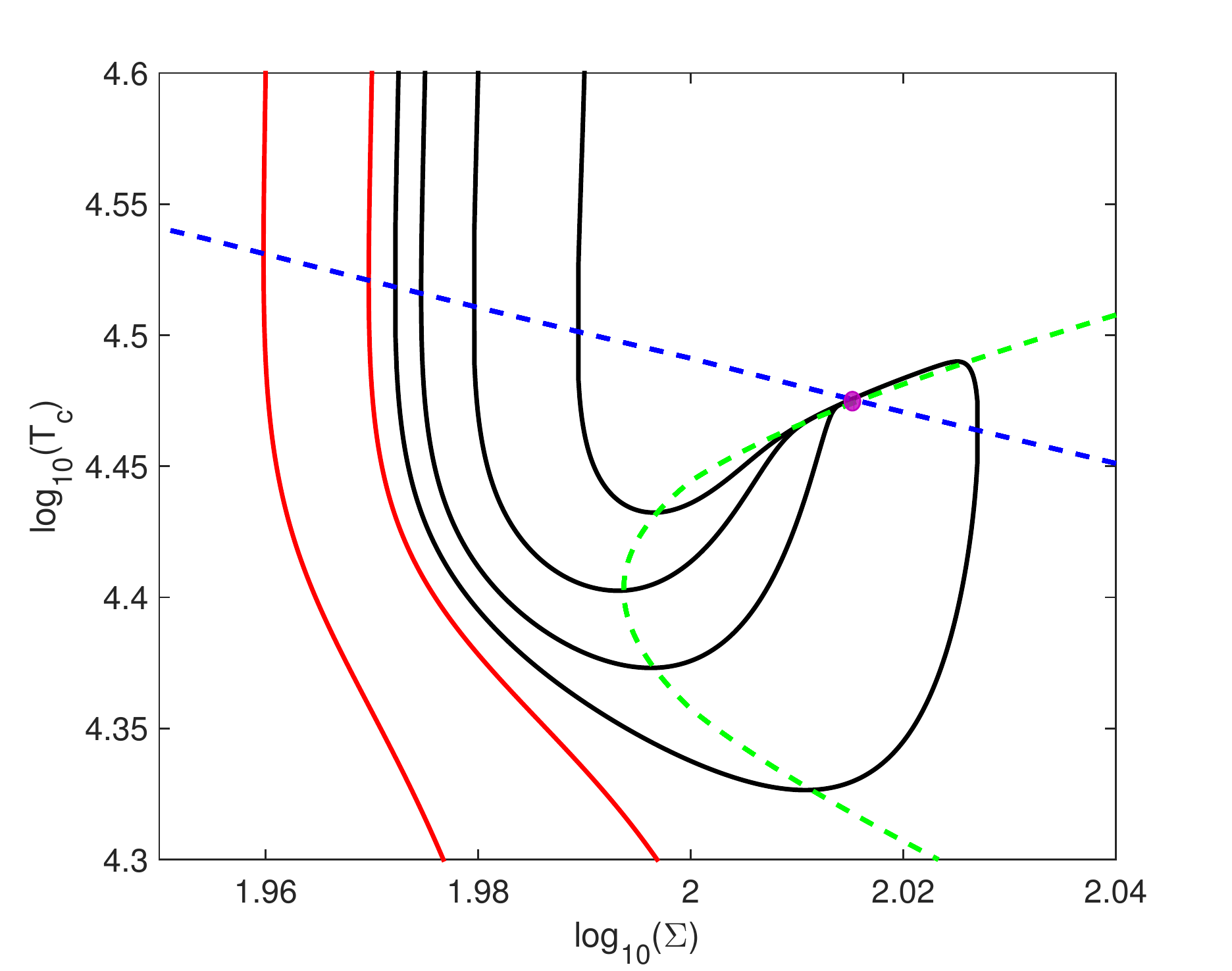}
\caption{ Right Panel: Deterministic trajectories close to the fixed point. Trajectories that undergo a full cycle are shown by the red lines and those that are immediately attracted to he fixed point are shown by the black. The fixed point is shown by the purple dot. }
\label{fig::contour_panel1}
\end{figure}

\subsubsection{Basins of attraction}

In a deterministic dynamical system the local basin of attraction of a
stable
fixed point is the region in phase space defined so that if a
trajectory begins in this region it will immediately return to the
fixed point. 
In Figure \ref{fig::contour_panel1} we plot deterministic trajectories through
 phase space started in the vicinity of the
fixed point. The figure indicates that there is a critical curve between regions where the
outburst trajectories lie and those that are attracted to the fixed
point --- somewhere between the two closest black and red curves.
This is the leftmost boundary of the local deterministic basin of attraction.
(Note that the whole phase space is within the \textit{global} basin
of attraction.)

\begin{figure*}
\includegraphics[width=8cm]{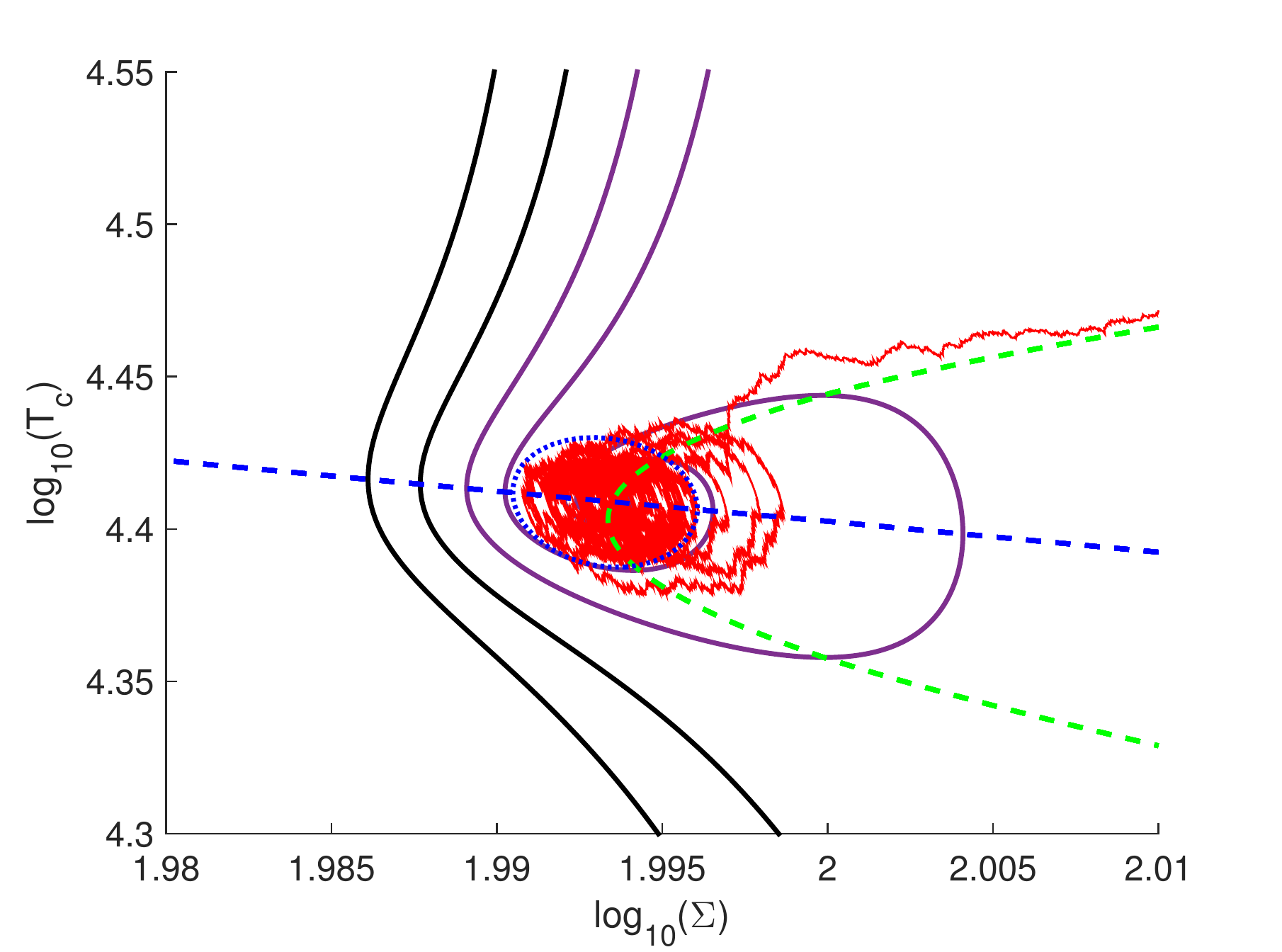}
\includegraphics[width=8cm]{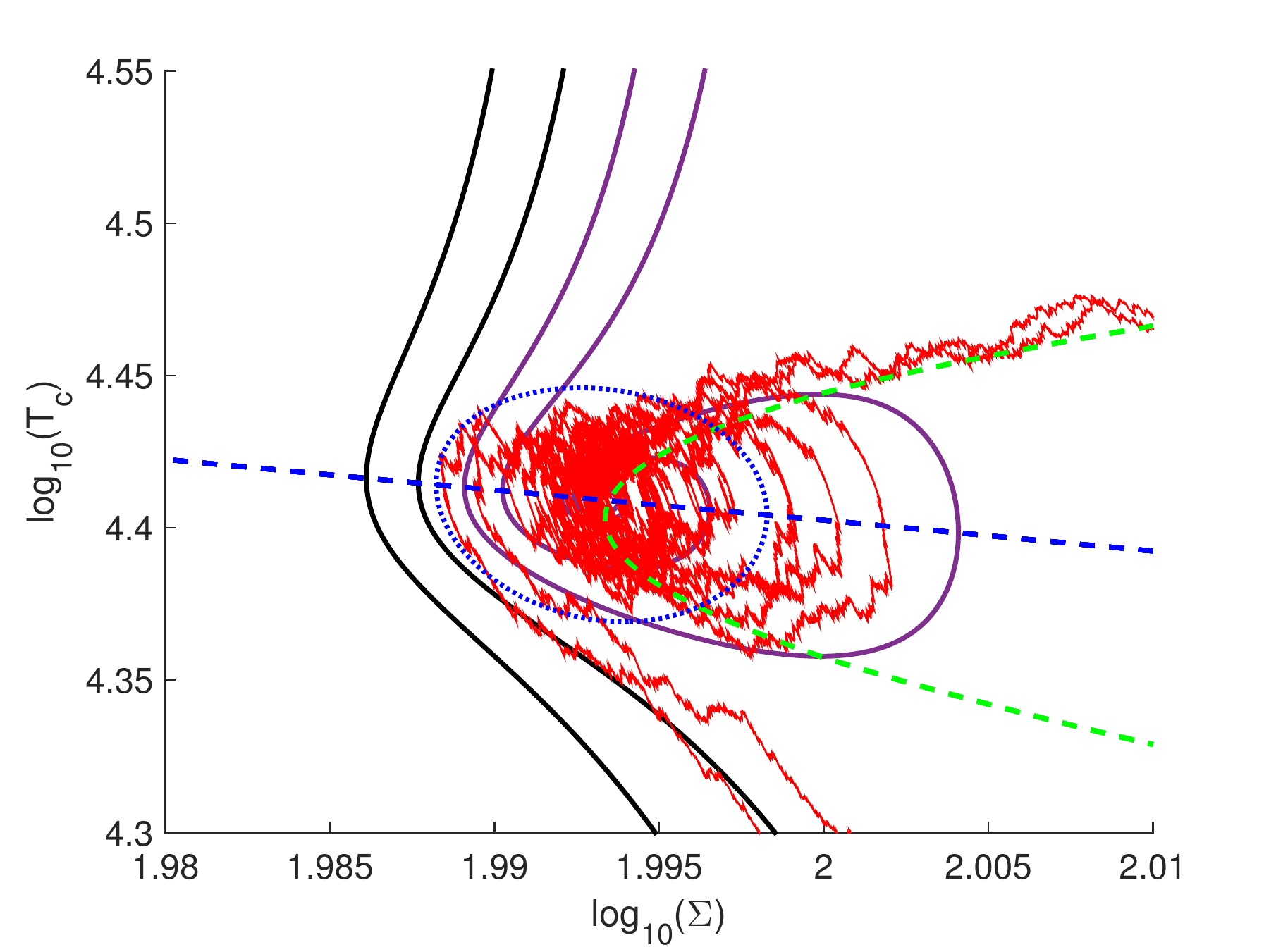}
\caption{Here $\beta(t)$ is a Wiener process with $b_0=0.5$ and
  $0.9$, and 
$\dot{M}=4\times10^{16}\,\text{g s}^{-1}$. Confidence ellipses of
$85\%$ are shown in dotted blue along with the nullclines in dashed 
green and blue. 
Deterministic trajectories
are shown by the black and purple lines, black indicates 
 a trajectory performing a limit cycle and purple 
indicates a solution attracted directly to the fixed point. Finally,
the red curve is the simulated trajectory.}
\label{fig::CE}
\end{figure*}

In a noisy system, such as ours, it is possible for fluctuations to
perturb a trajectory initially resting on the fixed point beyond the critical
boundary, at which point it will undergo an outburst. The fluctuation
must, however, be sufficiently large and/or the fixed point
sufficiently close to the boundary. 
For fluctuations written in terms of a Markov chain, as is the case
here, it is non-trivial to estimate the noise amplitude
that is necessary for excitation. However, if $\beta$ is assumed to be
white noise instead, then the problem is more analytically
tractable. In fact this has been tackled for the FHN model where
`confidence ellipses' can be calculated. 
A confidence ellipse of probability $p$ corresponds to an ellipse in
phase space enclosing the fixed point in which there is probability
$p$ of finding the solution at time $t$. In the Appendix we describe
how these may be calculated.

In Figure \ref{fig::CE}
two close-ups of the fixed point are shown. These include (a) deterministic
trajectories which help indicate the boundary of the basin of
attraction (solid black and purple), (b) the two nullclines (dashed blue
and green), (c) sample trajectories from a stochastic simulation
(red), and finally (d) 85\% confidence ellipses enclosing the fixed
points (dotted blue). The left panel corresponds to white noise of
amplitude $b_0=0.5$ and the right panel to more vigorous noise,
$b_0=0.9$. In the left panel the confidence ellipse is well within the
basin of attraction, and indeed the red trajectory remains trapped
around the fixed point. In the right panel the confidence ellipse is
larger, on account of the larger amplitude noise, and straddles the
basin's boundary. As a result, trajectories can undergo outbursts,
because the fluctuations aperiodically expel them from the fixed point's
basin of attraction. Indeed, we see here the red trajectory leave the
vicinity of the fixed point and disappear out of the figure.

\begin{figure*}
\includegraphics[width=15cm]{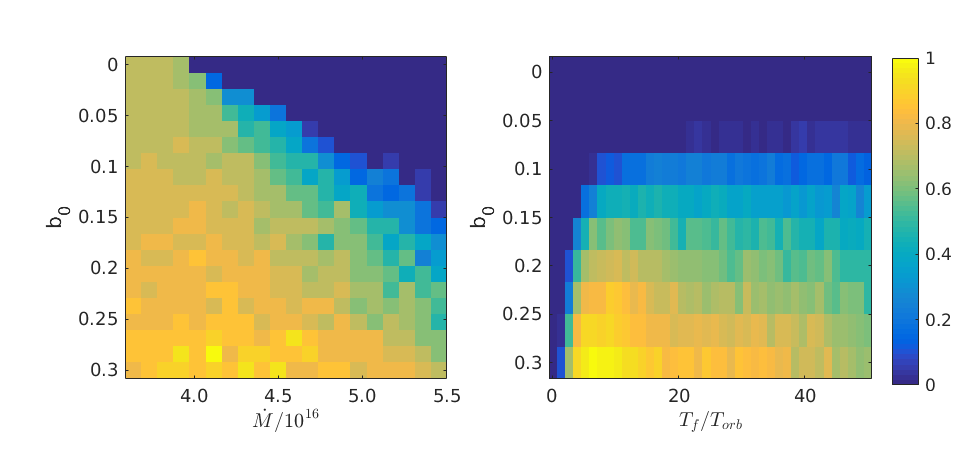}
\caption{ Both panels show the number of outbursts normalised by the
  maximum number of outbursts that occur during an interval of
  $5\times10^{7}s$ over the whole phase space shown using the one-zone
  model. Left panel (a): fluctuation amplitude and the mass accretion
  rate are both varied while $t_{f}/t_{orb}$ is fixed at $10$ . Right
  panel (b): 
fluctuation amplitude and timescale are varied while
 the mass accretion rate is kept fixed, $\dot{M}=5\times10^{16}$. }
\label{fig::contour_panel}
\end{figure*}

\subsubsection{Parameter dependence}

In this one zone model there are various parameters that may affect
the evolution of the disk annulus. Here we focus on $b_{0}$, $\dot{M}$
and the fluctuation timescale, $t_{f}$.

In the four panels of Figures \ref{fig::0DMdot} we show the
effect of changing $b_{0}$ and $\dot{M}$. The value of $\dot{M}$,
along with the radial location, determines the intersection of the two
nullclines, which in turn gives
the fixed point. There exists a critical mass transfer rate
$\dot{M}_{crit}$ (for this set-up $\approx 4\times10^{16}\text{g s}^{-1}$)
above which the fixed point is stable and below unstable. Obviously we
are interested in $\dot{M}>\dot{M}_{crit}$. Increasing $\dot{M}$
results in the fixed point moving to higher $T_{c}$ and $\Sigma$ and
hence moving it away from the bifurcation point. In Figure
\ref{fig::0DMdot} we see that this has the effect of decreasing the
outburst frequency by increasing the basin of attraction: the top left
panel exhibits fewer outbursts than the top right panel.
The amplitude of the fluctuations is given by $b_{0}$, increasing this
improves the chance of excitation, as evident in the bottom panels of
Fig.~\ref{fig::0DMdot}. For sufficiently small $b_{0}$ no
cycles occur because the fluctuations are insufficiently intense to perturb the
system into the outburst region. It is clear 
that a wide range of
outburst-limit cycles variation can be achieved - from U Gem (Figure
\ref{fig::0DMdot}b) to  outburst/dip pairs (panel d).
Note that sufficient noise amplitude permits U Gem type behaviour for
mass accretion rates somewhat above the critical value derived from
the classical theory.

\begin{figure*}
\includegraphics[width=18cm]{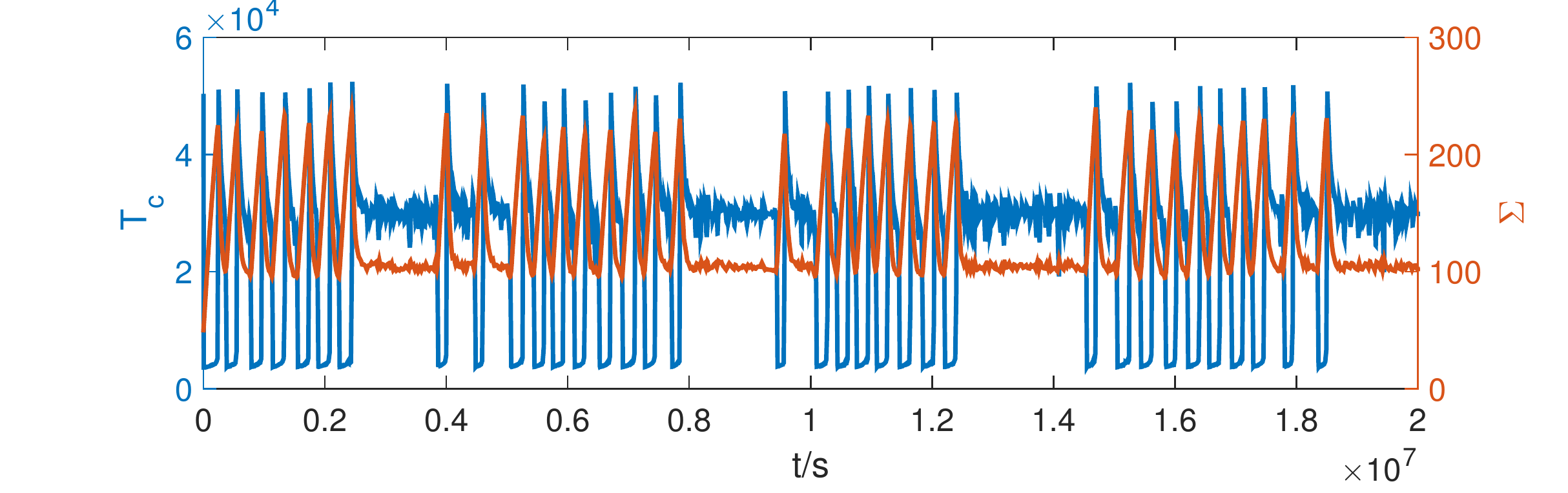}
\caption{Simulation including mass transfer variation,
  $\dot{M}=4.5\times10^{16}\text{g s}^{-1}$ with $\delta=0.1$,
  $T=5\times10^{6}s$ and $b_{0}=0.1$. }
\label{fig::run2}
\end{figure*}

In Figure \ref{fig::contour_panel}a we
investigate how the number of outbursts,
$N$, in a given duration of time depends on $b_{0}$ and $\dot{M}$. For
$\dot{M}<M_{crit}$ the system performs cycles even in the
deterministic case. In this region, increasing $b_{0}$ marginally
increases $N$. For larger $\dot{M}$, the deterministic system is
stable but there exists a critical $b_{0}$ above which the system is
stochastically excitable. This threshold is an increasing function of
$\dot{M}$. The reason for this is that as $\dot{M}$ increases, a
larger `kick' is necessary to perturb the system sufficiently out of
the local basin of attraction for the fixed point. If $b_{0}$ is
increased further, $N$ begins to plateau as a result of the finite
limit cycle period. The range in $b_{0}$, over which $N$ increases,
for a  given $\dot{M}$, increases with $\dot{M}$.

In all of the above we have kept the fluctuation timescale, $t_{f}$,
fixed at $10t_{orb}$. This choice is motivated by numerical
simulations of MRI turbulence but is not terribly well constrained and is
dependent on the physics included. In fact, fluctuations should occur over a
range of timescales. In Figure \ref{fig::contour_panel}b we explore
the effect of changing this timescale, fixing
$\dot{M}=5\times 10^{16}\text{g s}^{-1}$ but varying $b_{0}$. In a similar way to
Figure \ref{fig::contour_panel}a, we explore the $b_{0}$-$\dot{M}$
parameter space by counting the number of outbursts in a given time
interval. For $t_{f}<2t_{orb}$ there are no outbursts for the range of
$b_{0}$ considered, this is because the system has insufficient time
to adjust to the variation in $\alpha$, given that the characteristic
timescale for $\Sigma$ is $t_{\Sigma}\approx20t_{orb}$. For larger
$t_{f}$ the system has longer to adjust to the fluctuations but also
the probability of getting a large cumulative fluctuation in a given
time interval decreases. The timescale for which the number of counts
is maximum is determined by the competition between these two
trends. For $b_{0}>0.1$, the number of counts peaks at approximately
$t_{f}=8t_{orb}$ before decreasing with $t_{f}$. Unsurprisingly,  
the peak is at less than $t_{\Sigma}$ given that variations 
in $\alpha$ have an amplitude much less than $\alpha_{0}$.

\subsubsection{Mass transfer variability}

Slowly varying the mass transfer rate from the secondary  
translates the $\Sigma$-nullcline vertically and with 
it the position of the fixed point. 
Consider the deterministic situation. If at some instance during a
mass transfer cycle, the fixed point lies on the unstable manifold
then outbursts can occur (Figure \ref{fig::ODZCam}), if not, then
the system will remain in either the high or low state. 
Fluctuations will have limited impact on the
first situation other than marginally altering the outburst duration
and magnitude. In the second case, the fluctuations can cause
excitations and the dynamics are thus altered. 

The parameters in Figure \ref{fig::run2} are chosen such that the
deterministic model flips between a stable state and one only
just below the saddle node, and hence unstable. (However, we have
achieved similar behaviour with the range of $\dot{M}$ fully confined to the
stable branch.)
Here $\dot{M_{0}}=4.5\times 10^{16}\text{g s}^{-1}$.
 and we introduce a small periodic
mass transfer variation with $\delta=0.1$ and
$t_{I}=5\times10^{6}s\approx 60$ days. 
As with the classical model, a small and easily achievable
 variation in mass
transfer (Viallet and Hameury, 2008) is sufficient to produce clear
transitions between standstills and outburst trains when viscous
fluctuations are included. The inclusion of fluctuations has three
effects: it breaks the periodicity of the lightcurves, allowing
variation in the number of limit cycles in each outburst train; the
range of possible $\dot{M}$ that can undergo limit cycles is somewhat 
extended; and during standstills the system undergoes
isolated outburst/dip pairs. The modelled behaviour does indeed do a reasonable
job at reproducing most of the features of Z Cam outbursts.

\subsection{Summary}

Clearly the one-zone model is a vast simplification of a real
astrophysical disk. One drawback is its inability to
capture
interactions between adjacent annuli. In addition, even though we make a wide
parameter sweep, the noise amplitude and timescale are input
parameters. To make further progress
these inputs must be constrained. That being said, the one
zone model provides a qualitatively clear representation of a disk
annulus in the outer 
regions of a steady state disk and 
hence provides a simple means of studying 
the effect of viscous fluctuations on instability.

Allowing for the model's drawbacks, it is clear
that introducing stochastic fluctuations alters the
properties of the dynamical system. In particular, stochastic
variations in $\alpha$ can cause the excitation of a deterministically
stable system that is at a fixed point lying sufficiently close
to the global bifurcation point. This results in the system following a
periodic orbit before returning to its fixed point.  Once an annulus has been
excited the flux  through it will change, adjacent annuli must
 then respond and may themselves undergo transitions
 leading to the propagation of transition waves through the disk.

\section{Viscous global model}

The previous section shows that an isolated
disk annulus is excitable by stochastic variation in
$\alpha$. Obviously the local model is an idealisation:
while in a steady state the approximation is somewhat
justified, but as soon as the mass supply rate is
altered on a short timescale compared to the viscous timescale, or
there exists steep temperature gradients such as during the
propagation of a transition wave, it breaks
down. Additionally, radial energy flux to and from adjacent annuli may
stabilise a local thermal runaway before a transition wave can be
triggered. It is therefore necessary to check that the outburst
variability is not limited to local models but
can in fact result in global-disk limit cycles. In this section we take the
next natural  step by extending this idea into a global-disk
setting. We evolve the vertically integrated time dependent 1D thin 
disk equations \eqref{Eqn::massDiff} and \eqref{eqn::Energy} along
 with the supplementary equations \eqref{eqn::radFlux} -
 \eqref{Eqn::match} 
and \eqref{eqn::alpha}.

\begin{figure*}
\center
\includegraphics[width=15cm]{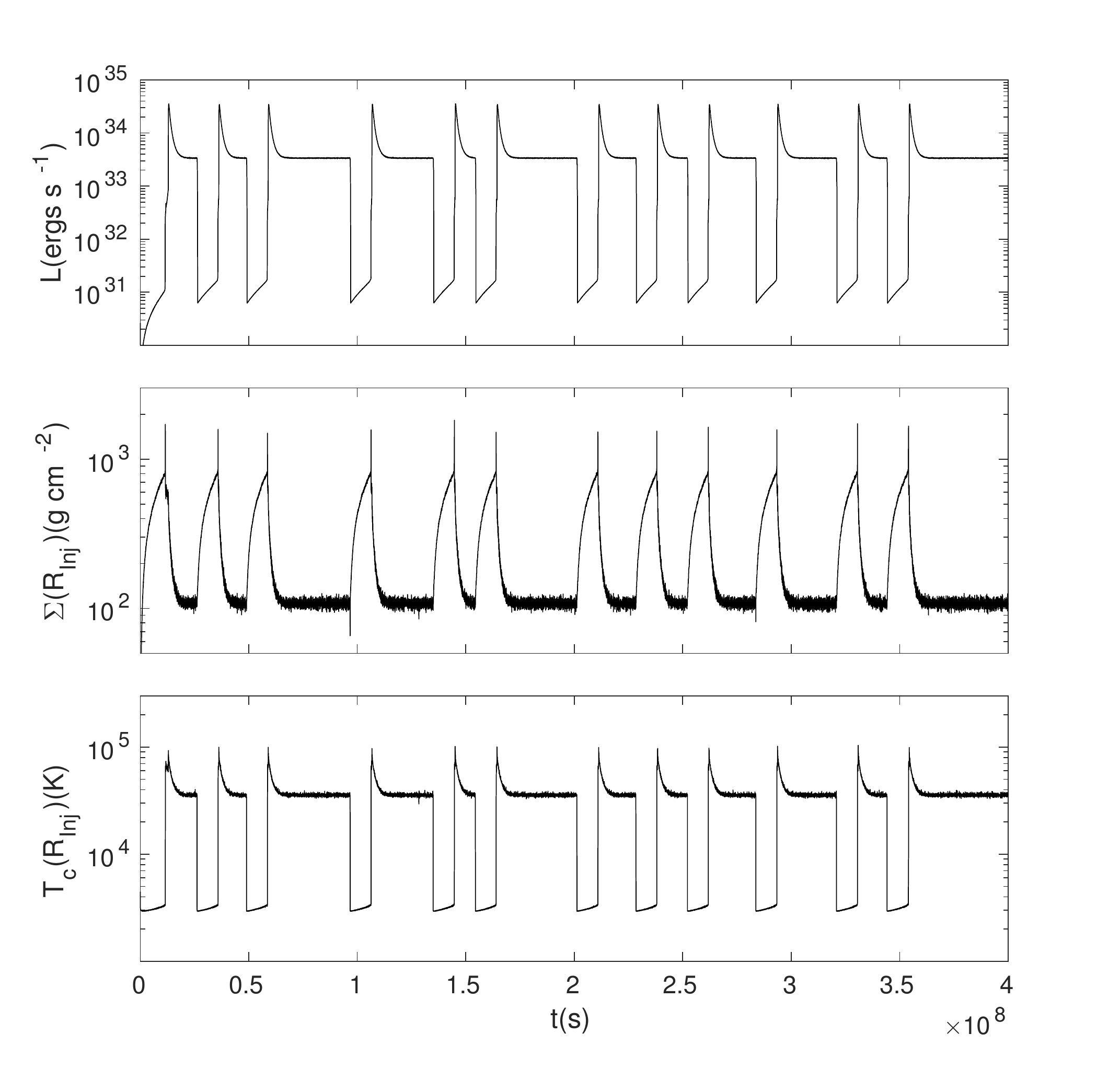}
\caption{Lightcurve from simulations with a mass injection rate of
  $1.1\times10^{17}\,\text{g s}^{-1}$, $b_{0}=0.125$ and $t_{f}=10$.
  Note the variable length of the standstills after each outburst,
  which differs from the (comparably regular) 
  viscous timescale of quiescence.}
\label{fig::D1LC}
\end{figure*}

\subsection{Viscosity}
Again we adopt the $\alpha$ formalism for the viscosity but also allow
for its spatial variation. It is well known that a multi-valued $\alpha$
is needed to explain the observed outburst durations and luminosity
magnitude changes in DNe (Smak 1984). We use
$\alpha_{0}=\alpha_{L}=0.01$ for the low temperature state and
$\alpha_{0}=\alpha_{H}=0.1$ for the high temperature state.
$\alpha$ has a stochastic component, $\beta$, which is a function of
radius as well as time. Therefore, Equation \eqref{Eqn::0Dfluctuations} becomes 
\begin{equation}
\alpha\left(t,r\right)=\alpha_{0}(t,r)\left[1+\beta\left(t ,r\right)\right]
\end{equation}   
where
\begin{equation}
\alpha_{0}(t,r)=
  \begin{cases}
\alpha_{L}, & \text{if}\ T_{c}(t,r)\leq10^4 \\
\alpha_{H}, & \text{if}\ T_{c}(t,r)>10^4.
  \end{cases}
\end{equation}
We assume that compressibility plays a central role in determining the
lengthscale of radial coherence in the effective viscosity and take
this scale to be the pressure scale height, $H(t,r)$ (Shakura \&
Sunyaev, 1973).  The timescale over which the fluctuations occur is 
 set to be a function of radius given by $t_{f}(r)=Ct_{orb}(r)$ where $C$ is a constant. 

In order to prevent abrupt discontinuities, 
we apply a slight spatial and temporal smoothing to the stochastic term.
Upon discretization,
\begin{equation}
\beta^{n}_{i}=b_{0}(1-c_{i})\left(0.6u^{n}_{i} +0.2u^{n}_{i-1}+0.2u^{n}_{i+1}\right) +c_{i}\beta^{k}_{i}
\end{equation}
where $u^{n}_{i}$ is the Markov process, given by Equation
\eqref{Eqn::0Dun}, at radial grid element $i$ at timestep $n$, and  
$k$ is the most recent timestep at which $u_{i}$ was
 updated. Note that $u^{n}_{i}$ is constant over radial intervals 
of order $H$ and is updated every $t_{f}$. The $c_{i}$'s are 
functions of time chosen such that a linear smoothing occurs over $t_{f}/10$. 
Physically, it is likely that there are neither instantaneous changes
nor abrupt discontinuities in the radial direction given the
properties of the large-scale MRI dynamo. Large-scale interactions (Guan et
al 2009),
 magnetic field linkages and compressive waves may also mediate spatial variations.

\subsection{Numerics, mass input, and boundary conditions}

We evolve Equations \eqref{Eqn::massDiff} and \eqref{eqn::Energy}
using a finite difference scheme on a fixed logarithmically
spaced grid of $200$ points. The advective term is solved using an
upwind method and we use multi-time-stepping for the radial radiative
energy transport term. The timestep is constrained by the most
stringent condition: the diffusive and advective CFL conditions
and a constraint coming from the local heating and cooling terms. The
outer disk edge is set at $3\times10^{10}\text{cm}$ and the inner edge at
$3\times10^{9}\text{cm}$. The boundary conditions are chosen for simplicity:
for the surface density we adopt $\Sigma(r_{in})=0$ for the inner
boundary, and the analytic solution to the thin-disc equations for the
outer boundary. The thermal energy equation is second-order in $r$ and
so two boundary conditions are necessary. However, as pointed out by
Hameury et al.\ (1998), higher order terms are of minimal importance
other than at transition fronts, and it is the local heating and
cooling that dictate what is going on. The boundary condition hence has  
minimal effect and we take $\partial T_{c}/\partial r=0$. 
Via the term $S$, we inject mass uniformly at a rate $\dot{M}$ in the range
 $2.4\times 10^{10}\text{cm}<r<2.5\times10^{10}\text{cm}$. 

Turbulent fluctuations have a coherence length set to the scale
height, $H$, as described earlier. 
When $H$ decreases below the grid size, we account for
averaging over a larger change in radius by introduce a factor
$H/\Delta R$ where $\Delta R$ is the radial bin size.
We also introduce a
buffer zones between  $r< 6\times 10^{9}\text{cm}$ and $r>2.8\times10^{10}\text{cm}$
to safeguard  against stochastic effects at the boundary.
Finally, we commence our simulations 
with a small initial surface density, $\Sigma=1\,\text{g cm}^{-2}$, 
and a temperature of $3000K$.  

\subsection{Luminosity}
Light curves are calculated using the bolometric luminosity defined through
\begin{equation}
L=2\int^{r_{out}}_{r_{in}}2\pi F_{s}rdr,
\end{equation}
allowing for comparison with observations. Note that the luminosity is
dominated by contributions from the inner regions due to the
functional  form of the emitted flux,  
and the radially decreasing temperature profile.

\begin{figure*}
\includegraphics[width=12cm]{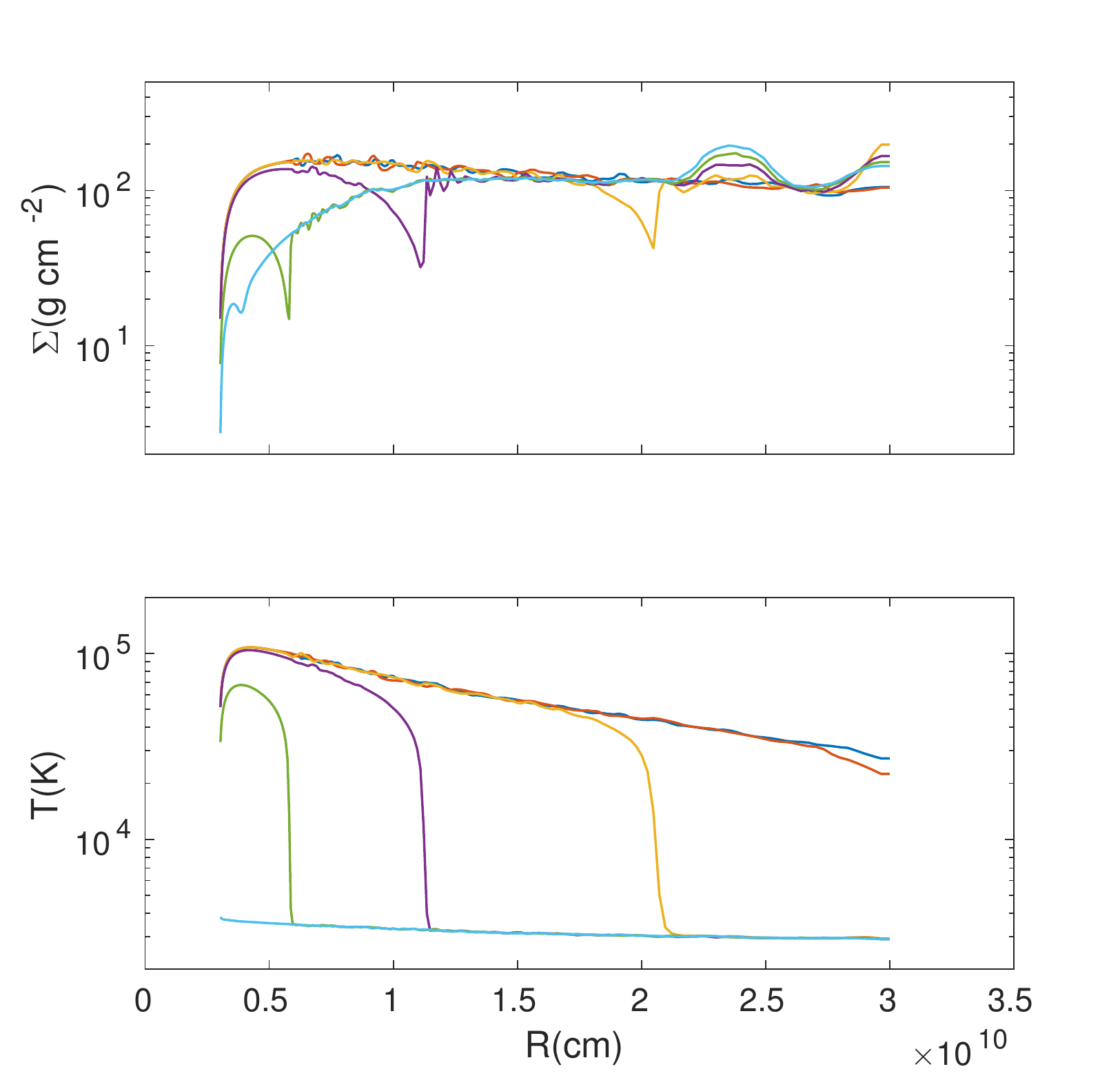}
\caption{Radial $\Sigma$ and $T_{c}$ profiles for the same simulation
  as in Figure \ref{fig::D1LC} during the excitation of a cooling
  wave. The profile are separated by $10^{5}$s starting with 
 the dark blue curve, proceeding through red, yellow, purple, green
 and ending with light blue.}
\label{fig::D1ZcamRadial}
\end{figure*}

\subsection{Results}

 Given that the outer
regions are more susceptible to downward transition we choose
parameters such that the whole disk is in the hot state but the
annuli near the outer edge are close to criticality --  
only marginally hotter than the upper bifurcation point. 
For our configuration this corresponds to
$\dot{M}\sim1.1\times10^{17}\text{g s}^{-1}$.
While the system is on the hot branch, the state that we are primarily
interested in, $H\sim 2\Delta R$ except at the inner 
boundary. During quiescence, typically $H\sim 0.5\Delta R$. 

For our first simulation we use $t_{f}/t_{orb}=10$, $b_0=0.1$, a
choice that is motivated by our parameter study in Section 3. The
resulting lightcurve is plotted in the top panel of Figure
\ref{fig::D1LC}. For the majority of the simulation, the disk
is in a high luminosity state but occasionally
transitions to a  quiescent
state two orders of magnitudes less luminous. 
The durations of quasi-steady periods range from
$\lesssim1\times10^{7}s$ to $\gtrsim4\times10^{7}s$, while the
quiescent phase  lasts for $\lesssim 1\times10^{7}s$.

In the middle and lower panels of Figure \ref{fig::D1LC} we show the
time evolution of $\Sigma$ and $T_{c}$ at the outermost injection
radius, $r=2.5\times10^{10}\text{cm}$. Qualitatively the behaviour at this
radius is similar to that in the one-zone model. The most noticeable
difference is the spike in the density that precedes the transition
to the hot state which is a result of the passage of the
pre-transition density wave (Papaloizou et al.~1983). 
The timescale amplitude is due to the  different imposed $\alpha$ in the low state.

Upon closer inspection, we find, as expected, that the cooling wave
originates close to the outer boundary where the disk is more
susceptible to downward transition, Figure
\ref{fig::D1ZcamRadial}. This feature is due to the shift of the
local $T$-nullcline towards
higher surface densities and the concurrent movement of the local
$\Sigma$-nullcline to lower surface densities. Once triggered
by a downward transition, a cooling wave, along with its preceding
density rarefaction wave (Papaloizou et al 1983), travels inward and
propagates through the full radial extent leaving the whole disk in
the low temperature/weakly accreting 
state with  a much  depleted surface density at small radii. 
As this wave has an interface of length $\sim H$ one may expect
appreciable pressure gradients and consequent wave launching, which
may be significant in the dynamics, though not captured by the 1D
diffusion model.

The minimal variation during the quasi-steady phases shown in the
lightcurve is a direct result of our inner buffer-zone. More
precisely, the emitted flux, $\propto T_{e}^{4}$, is a steep function
of $r$ for $r\gtrsim 4\times 10^{9}\text{cm}$ with its maximum inside our buffer region. 

The propagation distance of the cooling wave through the disk is
dependent on $\alpha_{H}$ and $\alpha_{L}$ (Smak 1984). Reduction of
this ratio from $\alpha_{H}/\alpha_{L}=10$ decreases the propagation
distance of the cooling wave through the disk and leads to reflection
of the cooling wave at some radius.  This shortens and reduces the amplitude of the
 luminosity drop, leading to sawtooth shaped lightcurves (Papaloizou et al 1983).

\subsubsection{Fluctuation amplitude}

\begin{figure*}
\center
\includegraphics[width=15cm]{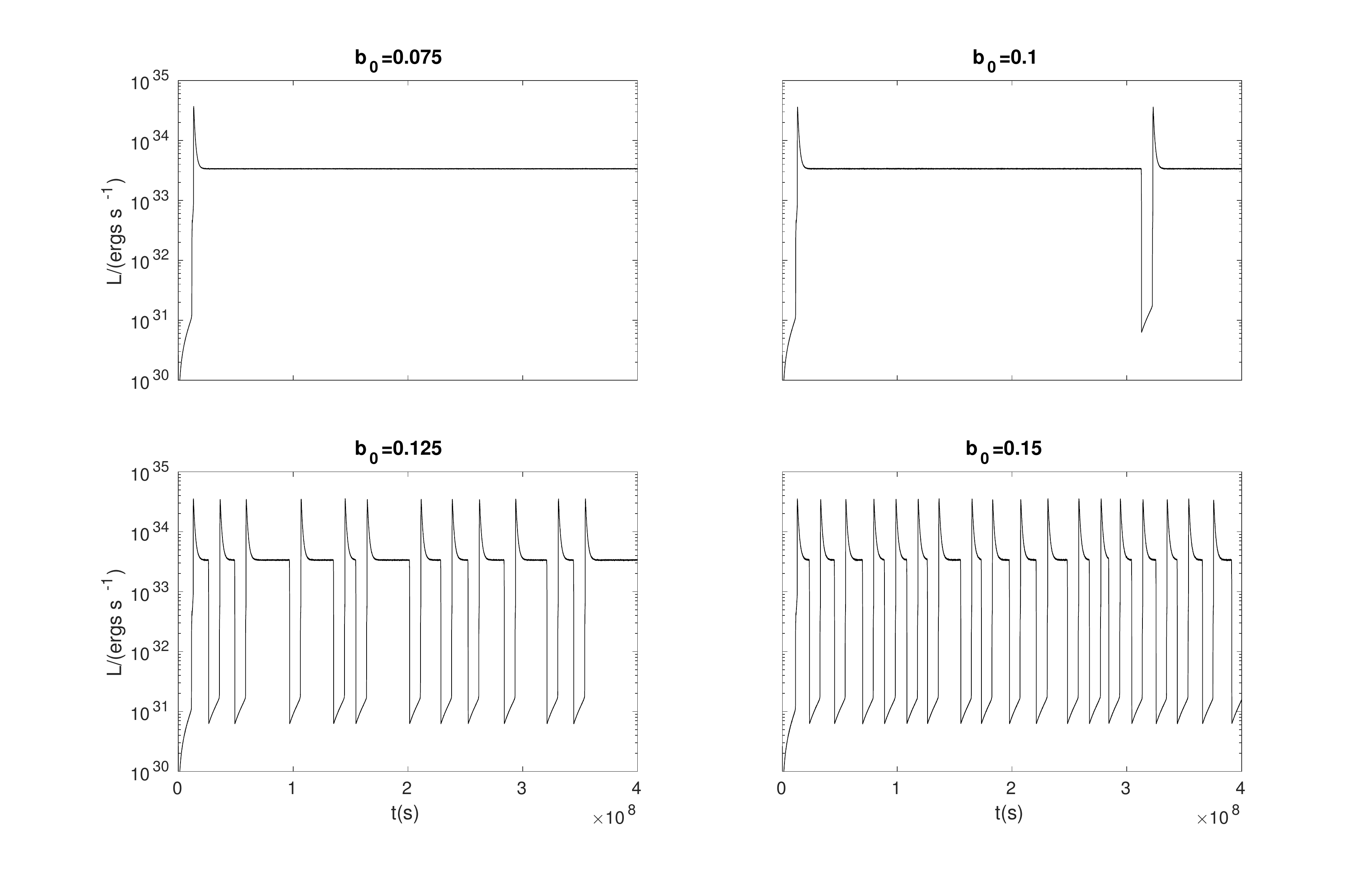}
\caption{Lightcurves from 1D simulations with
  $\dot{M}=1.1\times10^{17}\,\text{g s}^{-1}$ and $t_{f}/t_{orb}=10$.}
\label{fig::D1b0}
\end{figure*}

\begin{figure*}
\center
\includegraphics[width=15cm]{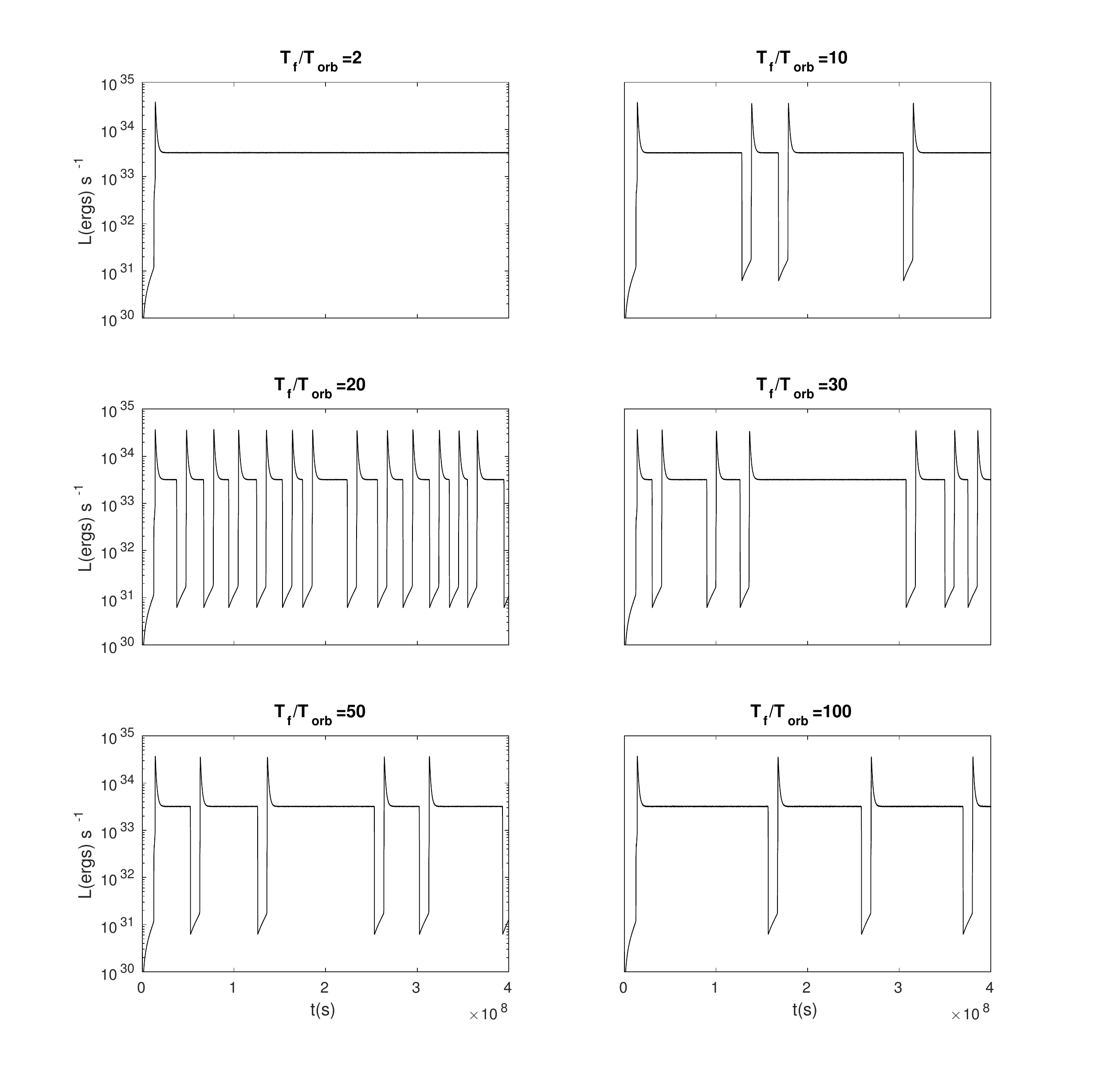}
\caption{Lightcurves from simulations with
  $\dot{M}=1.05\times10^{17}\,\text{g s}^{-1}$ and $b_{0}=0.075$ and various $t_{f}$.}
\label{fig::D1Tf}
\end{figure*}

To observe the effect of the fluctuation amplitude on excitability, we
fix $\dot{M}=1.1\times10^{17}\text{g s}^{-1}$ and $t_{f}=10t_{orb}$ but change
$b_{0}$. Light curves from simulations with various $b_{0}$ between
$0.075$ and $0.15$ are shown in Figure \ref{fig::D1b0}. As with the one-zone
model, the sensitivity to $b_{0}$ is apparent. For this
configuration, $b_{0}<0.1$ is unable to excite the system. However,
for $b_{0}\gtrsim0.15$ the disk undergoes near continuous limit cycles
more reminiscent of a U Gem system, but with some variability in
outburst length and occurance. This shows again that stochasticity can
transform a `stable' system into a U Gem system if the noise amplitude is
sufficiently large.
For a somewhat limited
intermediate band in $b_{0}$, the steady state durations are highly
non-uniform ranging between $3\times10^{6}s$ and $5\times10^{7}s$.    
We hence do not reproduce classical Z Cam behaviour: both the 
standstills and outburst trains are of too short a duration in
comparison
with observations.

\subsubsection{Fluctuation timescale}

In the one-zone model, fluctuations on timescales of $\lesssim t_{\Sigma}$
lead to the most regular outbursts. What is the timescale equivalent
to $t_{\Sigma}$ in the global disk simulations?
The one zone model states that
$t_{\Sigma}\sim (\Delta R)^{2}$. In a global disk the natural choice
for $\Delta R$ would be the turbulent lengthscale which we take to be
$\sim H$. However, in our
simulations we have a buffer zone at the outer edge and it is the outer
edge that goes unstable first, hence, a more apt choice is $\Delta R
\sim H+\Delta R_{\text{buff}}$. The surface density of this whole annulus
must evolve driven by fluctuations in density at the outermost
fluctuating orbit. With this we obtain $t_{\Sigma}\approx 30 t_{orb}$
for the diffusion timescale of interest. Therefore we expect the most
regular excitations when $t_{f}\lesssim 30 t_{orb}$. Note that this is
much less than the viscous timescale  if defined with respect to the whole disk.

In Figure \ref{fig::D1Tf} we plot lightcurves for a range of
$t_{f}/t_{orb}$ at fixed fluctuation amplitude and mass injection
rate. It is clear that if the fluctuation timescale is too short,
$t_{f}\ll t_{\Sigma}$, then excitations are unlikely/impossible. The
outer disk is unable to adapt to the fluctuations
sufficiently quickly. For $t_{f}$ closer to $t_{\Sigma}$, excitation
becomes more likely. The $t_{f}$ at which the disk is most excitable
appears to be $<t_{\Sigma}$ as was also the case in the one-zone
model. Increasing $t_{f}$ further leads to less frequent outbursts as
the disk annulus is able to adapt to the fluctuation on a shorter
timescale than $t_{f}$, resulting in idle time before the next
`kick'. Evidently long timescale fluctuations are unnecessary, and
furthermore non-optimal, for causing 
stochastic excitation; $t_{f}\lesssim H^{2}/\nu$ suffices.

     



\section{Discussion}

In this paper we study the stochastic excitation of DN accretion
disks caused by fluctuations arising from disk turbulence as
opposed to the more classical mass transfer variation. The systems of particular
interest here are the class of Z Cam variables. 
We intend this work to be a proof of concept and so adopt
simple local and 1D global time-dependent models. The radiative and
ionisation physics is approximated by a vertically
integrated cooling law (FLP), which is simple to
implement and physically motivated (though omits some of the detailed
physics).

First, we solve a one-zone system of stochastic differential
equations for the surface density and central temperature. 
DNe that would
otherwise remain indefinitely in a quasi-steady state can perform
limit cycles when the stress is allowed to be stochastic. The
duration of standstills prior to excitation depends strongly on the
fluctuation amplitude and the amount that the mass injection rate
exceeds the critical value determining
stability. Introducing a subcritical mass transfer rate
(insufficiently strong to cause transitions between cycles and steady
state behaviour) together with the fluctuations lead to trains of
outbursts separated by standstills. The standstills, however, are
somewhat shorter than observed, or the outburst trains not especially
long. It is, however, relatively easy to obtain
aperiodic singular outbursts.

Our second model is a 1D viscous time-dependent model, similar (but
somewhat simpler) than those
currently used (e.g.\ Cannizzo 1993a). 
We find that the outer regions
of a disk with a marginally supercritical mass supply rate can be
stochastically excited. Between excitation, the
standstills in the high state are found to last between several hundred
days to a few years. The transition from high to low states first
occurs at the outer boundary resulting in an inward propagating cooling
wave. When the average $\alpha$ in the low state is taken to be much
less than that of the high state the cooling wave can propagate
throughout the entire disk leading to  decreases in luminosity
 of $\sim$2 mag.
 For more comparable $\alpha$'s the cooling wave is reflected in the bulk of the disk. 

The results presented in this paper suggest that fluctuations in
accretion rate can trigger isolated drops from steady state to
quiescence and then back. The behaviour is reminiscent of the singular outburst/dip
pairs found in some Z Cams,
and possibly the stunted outbursts in Nova like variables (Honeycutt et al
1998, Honeycutt 2001). We believe it is natural to attribute
these events to disk
fluctuations. The mechanism can also generate repeated outbursts
amidst standstills, but either the standstills are too short (because
the disk turbulence can too easily perturb the equilibrium) or the
outburst trains too short to successfully match Z Cam light curves in general. 
Finally, sufficiently large viscous fluctuations can cause systems with
stable $\dot{M}$ to exhibit more classical U Gem type behaviour,
i.e. repeated outbursts with very few to no standstills.
Note that these results were obtained with only one, rather
simple, noise model and the fluctuations arising from MRI
turbulence may provide a more rich set of dynamics. 

We conclude that stress stochasticity on its own is not a viable
replacement
to mass transfer variability from the secondary in explaining
essential Z Cam phenomena (Meyer \& Meyer-Hofmeister
1983). This stochasticity, however, may still
provide an important ingredient in their dynamics, by 
adding an additional level of variability in the occurance and
duration of outburst trains and standstills, and may directly excite
aperiodic and isolated outburst/dip pairs. In fact, with more advanced
modelling it may be possible to constrain the fluctuation amplitude
via comparison with observed light curves.

\section*{Acknowledgements}

The authors would like to thank the reviewer, John Cannizzo, for a
set of helpful comments that improved the clarity of the paper.
The work also benefited from input from John Papaloizou and Pavel
Ivanov. HNL and JR are partially funded by STFC grants ST/L000636/1
and ST/K501906/1.

\bibliographystyle{apalike}
\bibliography{bib}

\begin{appendix}

\section{Confidence ellipses}

The introduction of a small amount of noise renders the FN model stochastically excitable,
leading to the `abberrant' firing of neurons (Linder et al
2004; Laing \& Lord 2010). Bashkirtseva \& Ryashko (2011) were able to
analytically determine noise thresholds that are necessary for
excitation. Here we give a brief summary of their technique and
then apply it to our local DN model.  

Consider the system of stochastic differential equations
\begin{equation}
\mathbf{\dot{x}}=\mathbf{f(x)}+b_0\mathsf{A}(\mathbf{x})\mathbf{\dot{w}},
\end{equation}
where $\mathbf{x}$ is an n-element vector, $\mathbf{f(x)}$ is
an n-vector function, $\mathbf{w(t)}$ is a n--dimensional Wiener
process, $\mathsf{A}(\mathbf{x})$ is a $n\times n$-matrix function,
and $b_0$ is the disturbance intensity. If there exists an attractor
when $b_0=0$, i.e. for the deterministic case, then a corresponding
stochastic attractor with stationary probability distribution function
$g(\mathbf{x},b_0)$ exists when $b_0\neq0$. The
\textit{probability distribution function} $g(\mathbf{x},b_0,t)$
is informally defined such that the probability that the solution is
within $d\mathbf{x}$ at time $t$ is given by
$g(\mathbf{x},b_0,t)d\mathbf{x}$. A \textit{stationary
  probability distribution} is then a probability distribution  
function that is time independent, $g(\mathbf{x},b_0,t)=g(\mathbf{x},b_0)$.

 This distribution is given by the Kolmogorov-Fokker-Planck equation
 which is difficult to solve in general. However, for small $b_0$ the distribution can be approximated by 
\begin{equation}
g(\mathbf{x},b_0)\approx K \text{exp}\left(-\frac{\Phi(\mathbf{x})}{b_0^{2}}\right).
\end{equation}
In the above expression $\Phi(\mathbf{x})$ is an appropriate
quasi-potential and $K$ a constant. 
If we assume that there is an exponentially stable
equilibrium, $\mathbf{\bar{x}}$, then a quadratic approximation for
the quasi-potential can be used
$\Phi(\mathbf{x})\approx\frac{1}{2}(\mathbf{x-\bar{x}})^{T}\mathsf{W}^{-1}(\mathbf{x-\bar{x}})$,
where $\mathsf{W}$, known as the stochastic sensitivity matrix, is a positive definite matrix that is uniquely determined by
\begin{align}
&\mathsf{FW}+\mathsf{WF}^{T}=-\mathsf{S}, \quad
\mathsf{F}=\frac{\partial \mathbf{f}}{\partial \mathbf{x}}(\mathbf{\bar{x}}), \\
&\mathsf{S}=\mathsf{GG}^{T},\quad
\mathsf{G}=\mathsf{A}(\mathbf{\bar{x}}). \quad
\end{align}
From this, a confidence ellipse surrounding the fixed point can be
computed from the following set of equations
\begin{equation}
(\mathbf{x-\bar{x}})^{T}\mathsf{W}^{-1}(\mathbf{x-\bar{x}})=2k^{2}b_0^{2},
\end{equation}
where $k=-\ln(1-p)$ and $p$ is the desired probability. A
\textit{confidence ellipse} 
of probability $p$ is defined as an ellipse around the fixed point for
 which there is a probability $p$ of the solution being within it at time $t$.  

With this information it is possible to estimate the amplitude of the
noise that is necessary to stochastically excite a system. 
Note that this treatment
is only possible when the fluctuations can be modelled as a white
noise, that is, the noise  has a correlation time that is 
negligible compared to the other timescales of the system.

In order to visualize the importance of the the perturbation amplitude
we calculate 85\% confidence ellipses and plot them in Figure
\ref{fig::CE}.
The left-hand figure shows that if the confidence ellipse is
contained within the local basin of the fixed point then outbursts are
very unlikely if not impossible. However, if the noise amplitude is
sufficiently large such that the confidence ellipse extends past the
critical curve separating the local basin of attraction from the limit
cycle phase space then stochastic excitation occurs. Essentially,
the question of whether stochastic excitability is possible or not is
determined by how likely it is for fluctuations to push  
the system over the critical curve and out of the local basin of attraction.

\end{appendix}

\end{document}